\documentclass[10pt,english]{elsart3p}   
   
\usepackage{commands}   
\usepackage{amsmath}   
\usepackage{amssymb}   
\usepackage[dvips]{graphicx}     
\usepackage[english]{babel}                                            
\usepackage[T1]{fontenc}   
\usepackage[dvipsnames]{xcolor}   
\usepackage{xspace}    
\usepackage{multirow}   
\usepackage{url}
\usepackage[pagewise,switch,displaymath]{lineno} 
   
\long\def\symbolfootnote[#1]#2{\begingroup%   
\def\thefootnote{\fnsymbol{footnote}}\footnote[#1]{#2}\endgroup}

\newcommand {\ensm}			{\ensuremath}   
\newcommand {\DD}[1]                    {\ensm{\mathinner{\Delta#1}}}   
\newcommand {\abs}[1]			{\ensm{\left\lvert{#1}\right\rvert}}   
\newcommand {\ve}[1]			{\ensm{\boldsymbol{#1}}}   
\newcommand {\defuni}[1]		{\ifmmode \mathrm{#1} \else ${#1}$ \fi}   
\newcommand {\um}[1]			{\defuni{\; #1}}   
\renewcommand {\sci}[2]                 {\mbox{\ensm{ #1 \! \cdot \! 10^{#2} }}}   
\newcommand {\scix}[1]                  {\mbox{\ensm{ 10^{#1} }}}   
   
\renewcommand {\degr}			{\ensuremath{^{\circ}}\xspace}

\newcommand {\pton}[1]			{\ensm{\left(  #1 \right)}}   
\newcommand {\pqua}[1]			{\ensm{\left[  #1 \right]}}

\newcommand {\minispazio}		{\;\;\;}   
\newcommand {\spazio}			{\;\;\;\;\;\;}   
\newcommand {\punto}			{\minispazio .}   
\newcommand {\virgola}			{\minispazio ,}

\newcommand {\beq}			{\begin{equation}}   
\newcommand {\eeq}			{\end{equation}}

\renewcommand{\deriv}[2]                  {\cfrac{\diffl{#1}}{\diffl{#2}}}

\renewcommand{\Fnumb}		{\ensuremath{F_\mathrm{n}}\xspace}   
\renewcommand{\Cherenkov}	{Cherenkov\xspace}   
\renewcommand{\FoV}		{FoV\xspace}   
\renewcommand{\Xmax}		{\ensuremath{X_{\mathrm{M}}}\xspace}   
\renewcommand{\Xris}		{\ensuremath{X_{\mathrm{R}}}\xspace}   
\renewcommand{\Nmax}		{\ensuremath{N_{\mathrm{M}}}\xspace}   
\renewcommand{\Gmax}		{\ensuremath{\gamma_{\mathrm{M}}}\xspace}   
   
\newcommand{\UHECP}		{{\cal UHECP}\xspace}   
\renewcommand{\EAS}		{{\cal EAS}\xspace}   
\renewcommand{\EUSO}		{{\cal EUSO}\xspace}   
\newcommand{\SEUSO}		{s{\cal EUSO}\xspace}   
   
\newcommand{\FoVAzimuthAngle}{\ensm{\phi}}   
   
\renewcommand{\FoVShowerAzimuth}	{\ensuremath{\psi_{a}}\xspace}   
\renewcommand{\Nmax}			{\ensuremath{N^\mathrm{ch}_\mathrm{M}}\xspace}

\begin{document}   
   
\begin{frontmatter}   
   
\title{The observation of Extensive Air Showers from an Earth-Orbiting Satellite}   
\maketitle   
   
\author[Genova]{M. Pallavicini},   
\author[Genova]{R. Pesce\corauthref{cor1}},   
\author[Genova]{A. Petrolini},   
\author[Genova]{A. Thea\thanksref{AleAt}}   
   
\corauth[cor1]{Corresponding author. Address: Dipartimento di Fisica dell'Universit\`a di Genova and INFN, via Dodecaneso 33, I-16146 Genova, Italy. E-mail: Roberto.Pesce@ge.infn.it}   
   
\address[Genova]{Dipartimento di Fisica dell'Universit\`a di Genova and INFN, via Dodecaneso 33, I-16146 Genova, Italy.}   

\thanks[AleAt]{Now at Eidgen\"ossische Technische Hochschule (ETH), Z\"urich.}   
   
   
\begin{abstract}   
   
We review the main issues that are relevant for the observation of Extensive Air
Showers from an Earth-Orbiting Satellite.  Extensive Air Showers are produced by
the interaction of Ultra-High Energy Cosmic Particles with the atmosphere and
can be observed by an orbiting telescope detecting the air scintillation light.

We provide the main analytical formulas and semi-analytical results needed to
optimize the design of a suitable telescope and estimate the best-expected
performance and the minimal necessary requirements for the observation.

While we have in mind an \EUSO-like general-purpose experiment, the results presented in this
paper are useful for any kind of space-based experiment.
   
\end{abstract}   
   
\end{frontmatter}

%\linenumbers
%--------------------------------------------------------------------------------   
\section{Introduction}   
\label{sec:Intro}   
%--------------------------------------------------------------------------------   

Ultra-High Energy Cosmic Particles (\UHECP), with energies in excess of   
$E\approx\scix{19}\um{eV}$, hit the Earth with an extremely low flux of about   
\mbox{one $ \um{particle \cdot km^{-2}\cdot sr^{-1}\cdot millennium^{-1}}$},    
for energies above \mbox{$ E\approx\scix{20}\um{eV} $}~\cite{Abraham:2008ru}.   
   
The observation of \UHECP and the interpretation of the related   
phenomenology is one of the most interesting topics of contemporary   
High-Energy Astro-Particle Physics.  Direct detection is impossible at   
these energies, due to the exceedingly low flux, but \UHECP can be   
detected by observing the Extensive Air Showers (\EAS) produced by the   
interaction of the primary particle with the Earth atmosphere.   
For recent reviews on these topics see~\cite{Zavrtanik} and references therein.   
   
Two ground-based experiments, the Pierre Auger Observatory (PAO)~\cite{Watson:2008zzb} and
the Telescope Array (TA)~\cite{Matthews:2011zz}, are
currently taking data and hope to provide a clear   
understanding of many important
topics in the next few years~\cite{Abraham:2008ru,Abraham:2007si,Abreu:2010zzj}. 
Using an hybrid detection technique, PAO and TA will improve on the results collected 
in the past by HiRes~\cite{Abbasi:2007sv} and AGASA~\cite{Takeda:1998ps}.
However, it is likely that the next generation of  
experiments, after PAO and TA, will use an Earth-orbiting satellite, in   
order to increase the event statistics, by exploiting the huge   
instantaneous geometrical aperture potentially available to such an experiment.   
   
The aim of this paper is to discuss and summarize a few key issues   
relevant to the design and optimization of space-based experiments for   
the observation of \UHECP. Analytical formulas and semi-analytical results   
will be presented and discussed, in order to define the main parameters   
for the observation of \EAS from space. We also summarize and update the most important results   
presented in~\cite{Pallavicini:2008wy}, where more details can be found.   
A preliminary study was carried on in~\cite{arisaka},   
which was the starting point for some of the results presented in this paper.   
Although the results presented in this paper refer to an \EUSO-like general-purpose experiment, most of them
can be applied or easily re-adapted to any space-based experiment for the observation
of \EAS.   

The design of a telescope for the detection of \EAS from space is a very
challenging task, because of the low expected rate of events, the faint signal
received from any \EAS, the harsh space environment and the tight technical
constraints imposed on a space experiment (mainly on mass, power, volume and
telemetry).  The engineering is very complex and the design has a strong impact
on the scientific performance. A careful design optimization is therefore
mandatory.
   
Some of the assumptions of this paper are somewhat optimistic with respect to the
real conditions of the experiment, in order to be able to derive 
analytical formulas and semi-analytical results.  
We will neglect, for instance, the effect of clouds, of Mie
scattering, of multiple scattering and of different types of background other
than the random night-glow background; imperfections of a real
apparatus will not be taken into account as well.  In fact none of these effects
can be easily described in a semi-analytical way.  More precise and detailed
results, based on more realistic conditions, can be only obtained by means of a
full Monte-Carlo simulation of a specific experimental design: see the complementary
paper~\cite{bib:esaf}.  Therefore the results of this paper are the minimal
necessary requirements for the observation, but possibly not sufficient ones.
   
The results presented in this paper are a basic input to the detailed
optimization and design of the telescope. These results cannot replace a full
Monte-Carlo simulation, for detailed studies of any specific experiment. However
they are extremely useful for a basic understanding, for defining a baseline
design and for a rough cross-check of the results of
detailed Monte-Carlo simulations. Therefore they provide a valuable starting
point preliminary to any full Monte-Carlo simulation, as the latter requires a
well-defined design.  

Most of the work presented in this paper is the result of the   
development of the \EUSO experiment~\cite{EUSO}, a mission   
of the European Space Agency (ESA), which successfully completed its   
phase A study in 2004 and was frozen due to programmatic and   
financial reasons.  
The \EUSO Collaboration developed a full Monte-Carlo simulation,
ESAF~\cite{bib:esaf}, generating detailed prediction for an \EUSO-like telescope.

The optimization of the design of such a challenging experiment requires to collect as many as   
possible preliminary information. It is the opinion of the authors that
a number of preliminary and preparatory steps are mandatory~\cite{bib:petrolini1}.   
The most important one is a detailed characterization of the background via a suitable   
micro-satellite mission~\cite{spacepart06}, which might also test some of the technological issues.

The outline of the paper is as follows: 
the science case is briefly summarized in Section~\ref{sec:SciCase};
the observational approach is presented in Section~\ref{sec:Approach};    
the scientific requirements and the requirements derived for the experimental apparatus are presented in Section~\ref{sec:SciObj};
the assumptions are summarized in Section~\ref{sec:GenAss};
the design and optimization of an \EUSO-like experiment are discussed in Section~\ref{sec:Results}.

%--------------------------------------------------------------------------------   
\section{The scientific case}   
\label{sec:SciCase}   
%--------------------------------------------------------------------------------

The present knowledge of the the physics of \UHECP is still largely incomplete.  

The sources and the identity of the primary particles are not known.
Whether there is an end or not of the high-energy spectrum is not known either.
The sources are most probably extragalactic ones and relatively nearby, but they
have not been identified, yet~\cite{Abraham:2007si,Abreu:2010zzj}.
The existence of a cut-off in the energy spectrum seems to be
confirmed~\cite{Abraham:2008ru}, but this is in agreement with a
composition of the primary particles mostly consisting of protons, while data at
lower energies show that the mass of the primary particles
increases with the energy~\cite{Abraham:2007si}.

PAO and TA are collecting a large statistics of \UHECP and starting to answer
some of the open questions.  
However, a more complete understanding would require a
systematic study of \UHECP at energies above $ \approx 55 \um{EeV} $, where
ground-based experiments suffer from the limited statistics. In fact, at these high energies
there exist only a few measurements of the longitudinal profile of \EAS,
since the PAO detects and measures the longitudinal profile of about two
events per year, over a total of about twenty-five events observed per year.  
With this limited
statistics it is currently impossible to identify the sources of \UHECP and to
determine the primary particle identity.

A more detailed discussion of the scientific case for an \UHECP space experiment
can be found in~\cite{petrosanta}.

%--------------------------------------------------------------------------------   
\section{The observational approach (AirWatch)}   
\label{sec:Approach}   
%--------------------------------------------------------------------------------   

John Linsley, in 1982, first suggested~\cite{linsley}
that the Earth atmosphere at night,    
viewed from space, can act as a huge calorimeter for remotely observing \EAS (SOCRAS).   
Since then a number of proposals and studies were carried on, including the    
OWL~\cite{owl} (Orbiting Wide-angle Light-collectors) project,    
the TUS/KLYPVE~\cite{tus} project and \EUSO~\cite{EUSO}.    

In more recent times, the JEM-EUSO~\cite{jemeuso} Collaboration aims to propose
again the \EUSO concept on the International Space Station.
The \SEUSO
mission~\cite{supereuso,petrosanta,petrolini2}, for a challenging
next-generation experiment, was proposed to the ESA Cosmic Vision program
2015-2025~\cite{cosmicvision} and has been recommended by ESA for technological
developments.  
   
A telescope on an Earth-orbiting satellite, observing down the Earth at night
\footnote{The term \emph{space-based} will be used in this paper to refer to
  such an experiment, as opposed to the term \emph{ground-based}.},    
can detect the near-UV air scintillation light isotropically   
produced during the \EAS development in the atmosphere by the interaction of   
the \EAS secondary particles with the air molecules. 
The measurement of the isotropic air
scintillation light traces the development profile of the \EAS. 
Additional information can be gathered    
by observing the \Cherenkov light diffusely reflected by the Earth surface    
(by land, sea or clouds). The Earth atmosphere plays the role of a gigantic passive 
calorimeter, continuously changing and only usable at night.    
In fact the emitted light is proportional to the energy of the \EAS.
   
This approach is complementary to ground-based observations.  In fact any \EAS   
develops close to the Earth surface and, thanks to the large 
distance from the \EAS, a large Field of View (\FoV) space-based telescope   
can observe a large atmospheric target.
However, due to the large distance from the \EAS, the light signal is definitely
much fainter 
than the light signal observed by a ground-based experiment for the same \EAS.
Space-based observation is therefore   
best suited for observing low fluxes of high-energy \UHECP.  

The required telescope is an Earth-watching    
large aperture, large \FoV, fast and highly pixelized    
digital camera for detecting a moving spot of near-UV single photons    
superimposed on a large background, designed to operate in space five years, at least.   
   
The \EAS is seen as a quasi point-like image moving on the focal surface of the telescope with a   
direction and an angular velocity depending on the \EAS primary direction with respect to the   
line-of-sight.    
The space-time and intensity characteristics of the \EAS signal is used to distinguish the \EAS from
various types of background (see Section~\ref{subsec:NoiseBackground}),    
because the latter, typically, have a different space-time development.   
   
\begin{figure}[htbp]   
\begin{center}   
	\includegraphics[width=0.46\textwidth]{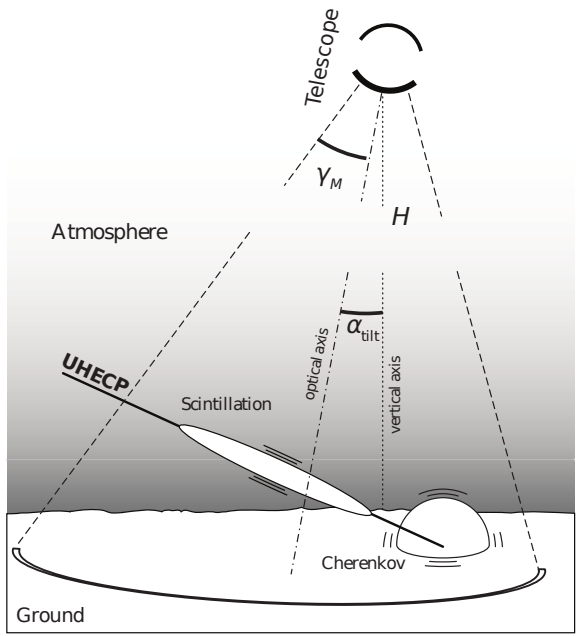}   
	\caption{AirWatch observational approach: $H$ is the   
	orbital height, $\Gmax$ is the \FoV half-angle and   
	\TiltAngle\ is the tilt angle between the optical axis and the   
	local nadir.}   
	\label{fig:approach}   
\end{center}   
\end{figure}

The observational approach is schematically shown in Figure~\ref{fig:approach}.   
   
\begin{figure*}[htb]   
\begin{center}   
	\begin{tabular}{cc}
	\includegraphics[width=0.46\textwidth]{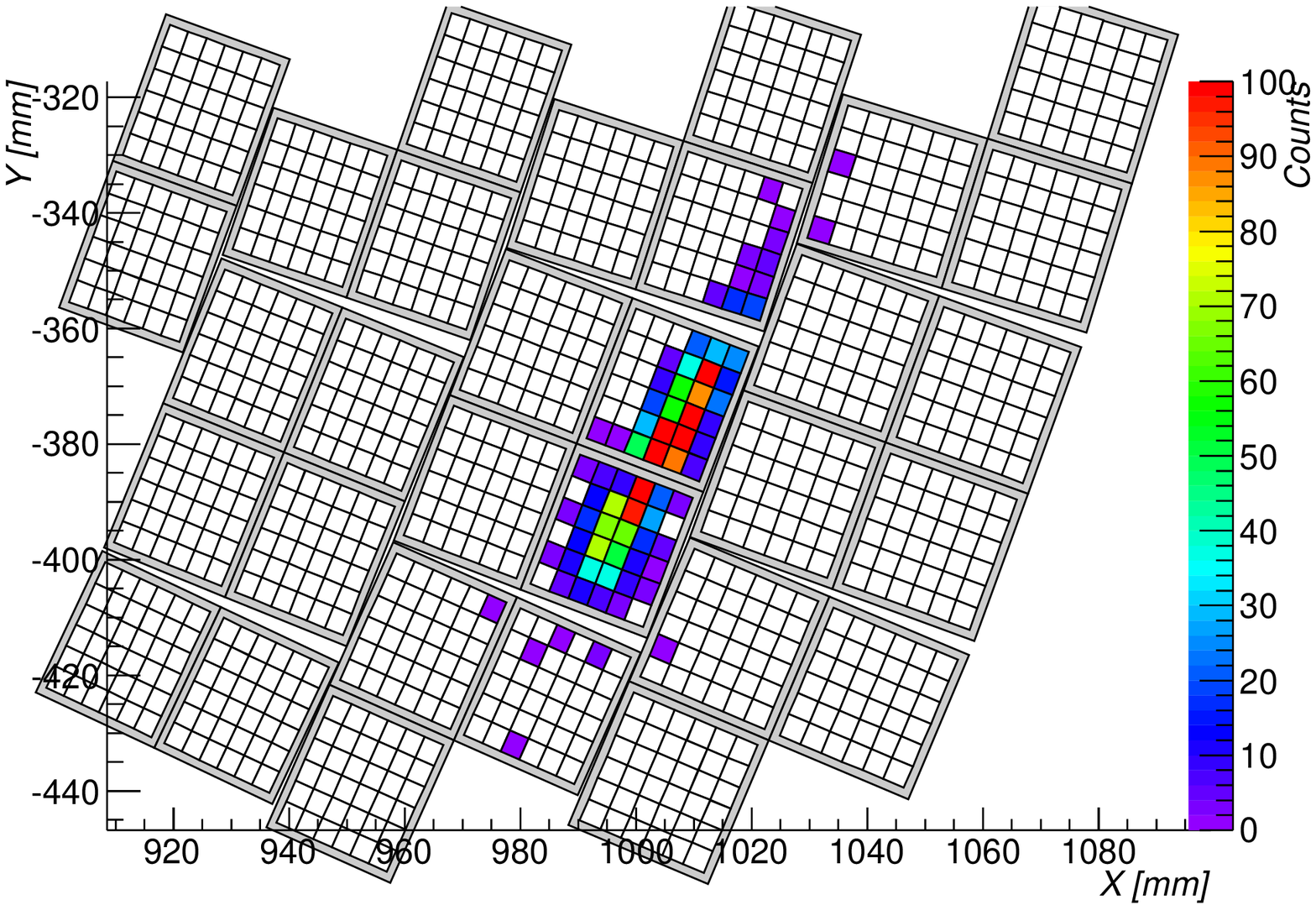} &
	\includegraphics[width=0.46\textwidth]{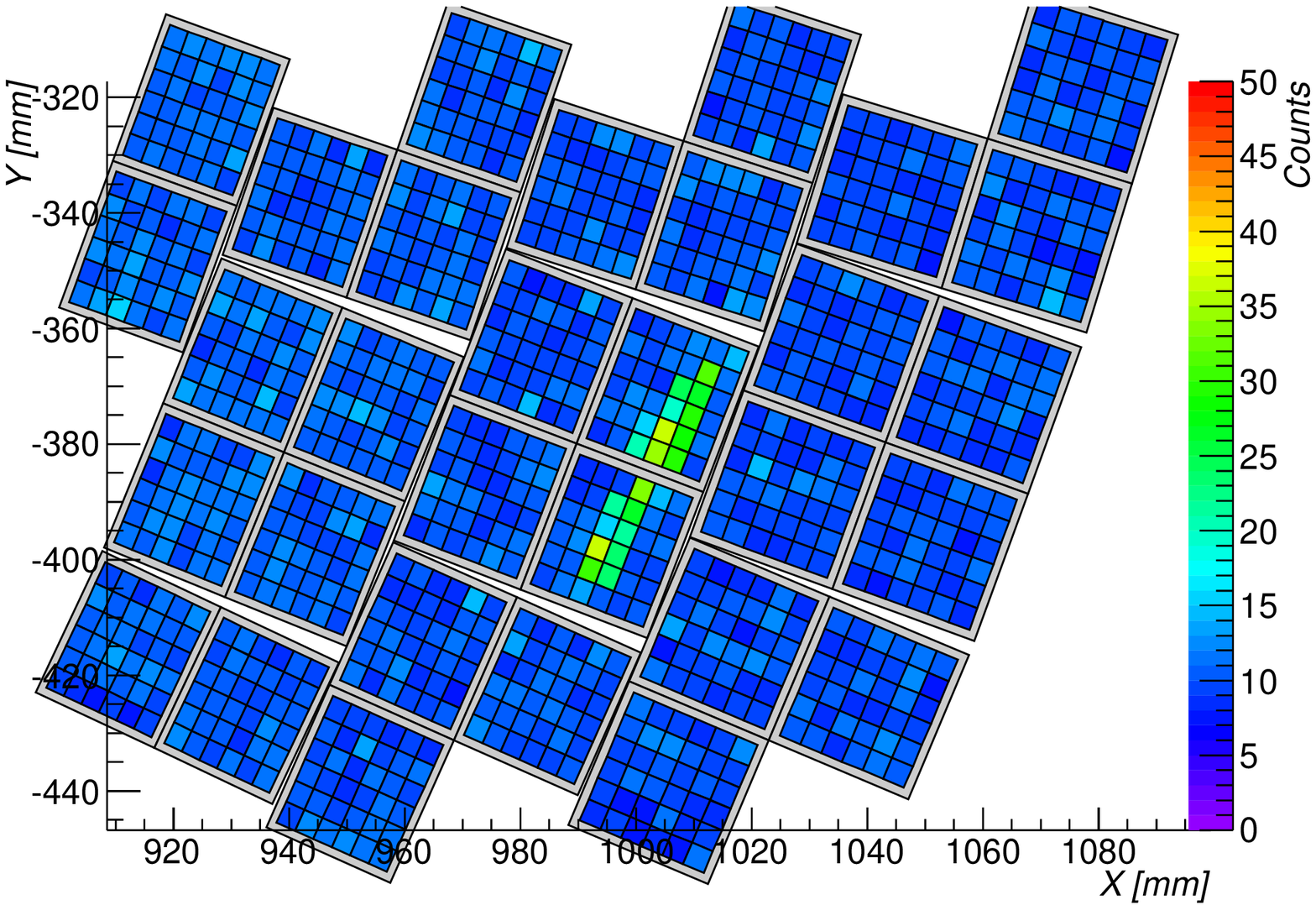} \\
	(a) & (b)\\
	\end{tabular}

\caption{Image of a typical simulated $ E= 10^{20} \um{eV}$ proton \EAS as seen on the focal
          surface of a typical $D=4\um{m}$ diameter space-based telescope at an
          altitude of 400~km above the sea level; the \EAS zenith angle is
          45\degr. 
          (a) EAS signal integrated in time (158~$\mu$s); the pixels without signal are in white and the PMT border is drawn in grey; 
          (b) the background is superimposed, but only the maximum number of
            counts in a pixel per time unit is shown. 
            The images are produced with ESAF~\cite{bib:esaf}.}   
	\label{fig:EASFS}   
\end{center}   
\end{figure*}

\begin{figure}[tb]   
\begin{center} 
\begin{tabular}{c}
	\includegraphics[width=0.385\textwidth]{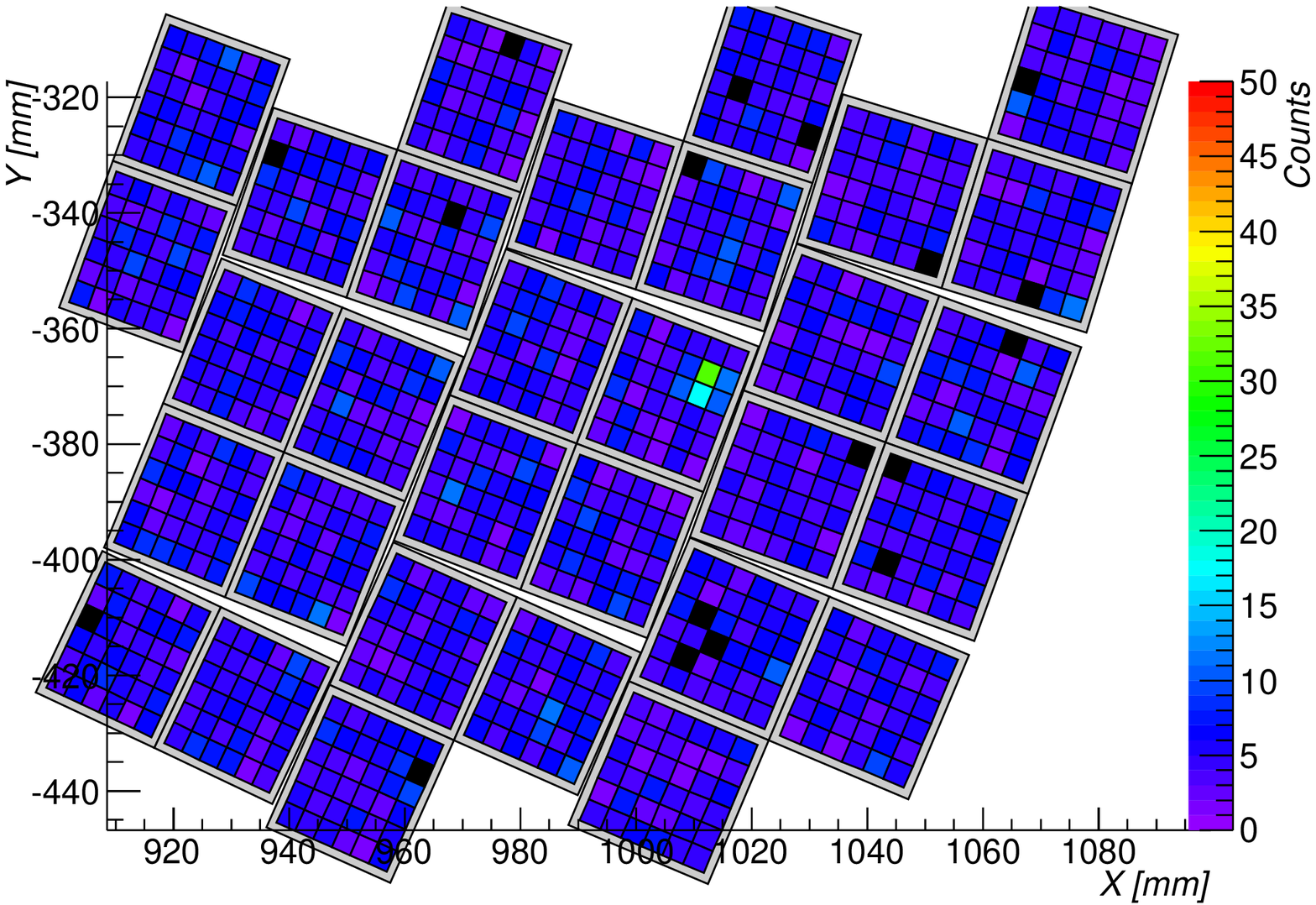} \\ 
        \includegraphics[width=0.385\textwidth]{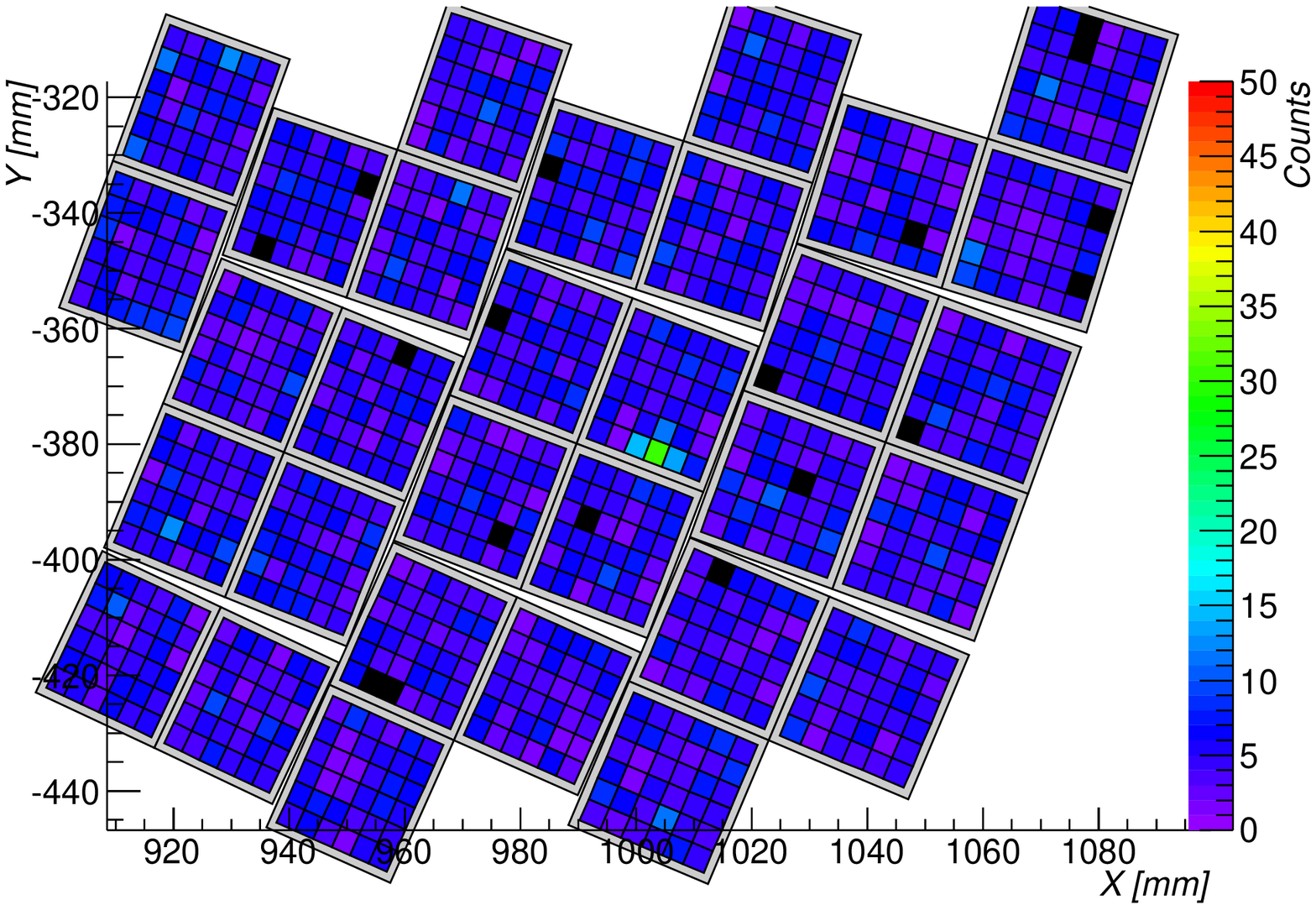} \\
	\includegraphics[width=0.385\textwidth]{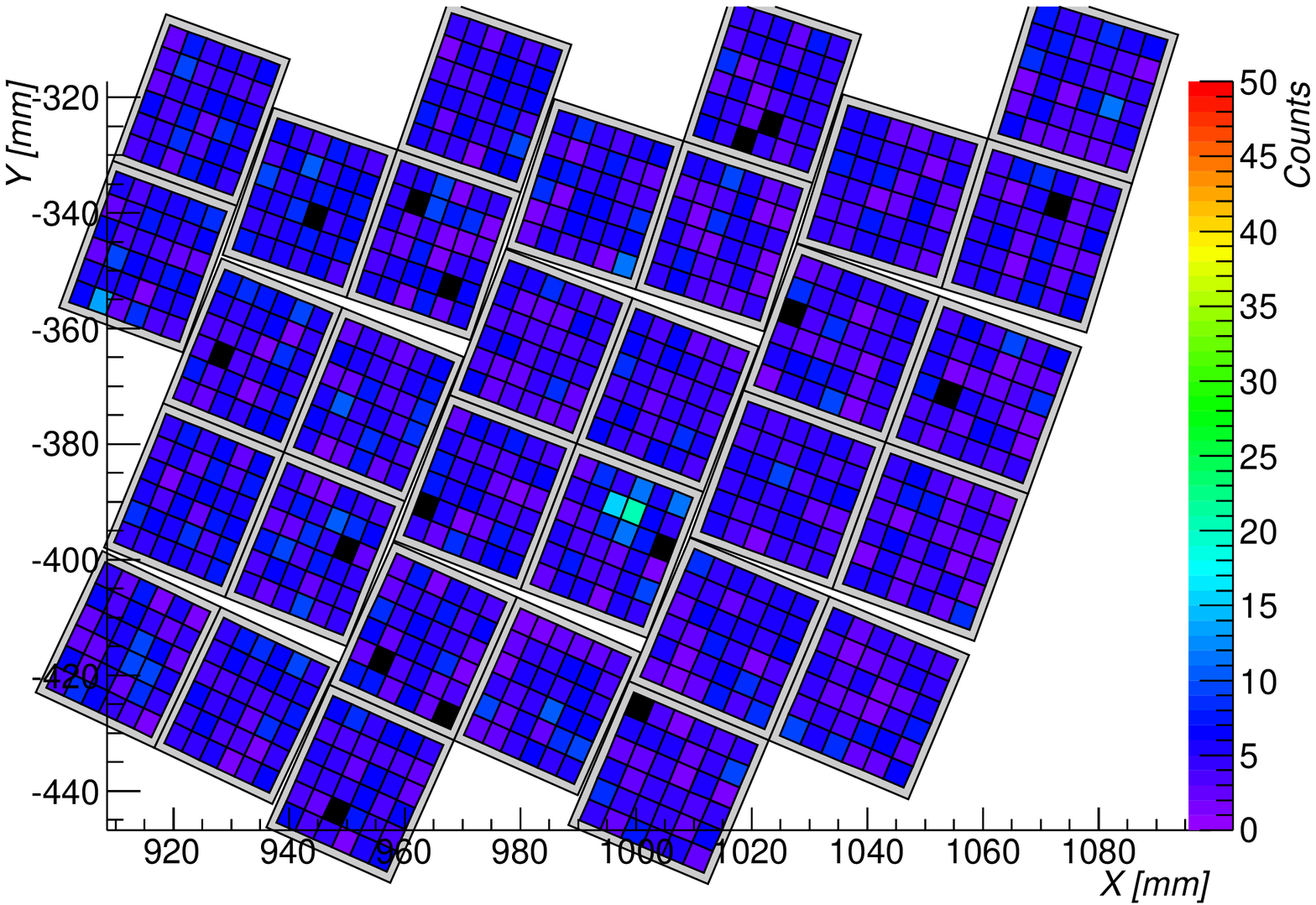} \\ 
        \includegraphics[width=0.385\textwidth]{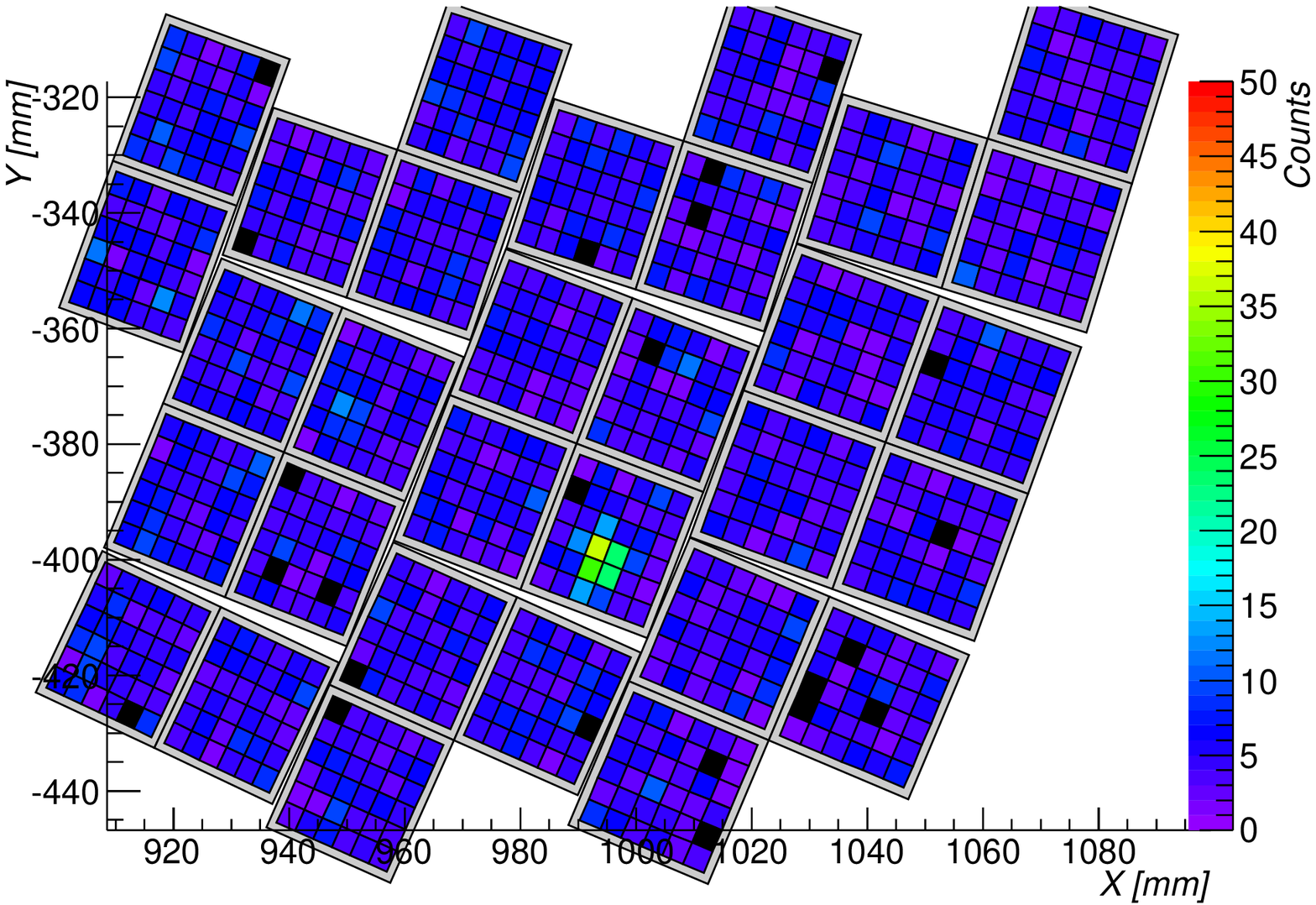} \\	
	\end{tabular}
\caption{Image of the same EAS of Figure~\ref{fig:EASFS} with   
          the background simulated. From top to bottom, the image at four different times (70~$\mu$s, 100~$\mu$s, 115~$\mu$s, 133~$\mu$s)
          is shown. The off pixels are in black and the PMT border is drawn in grey. The images are produced with ESAF~\cite{bib:esaf}. } 
\label{fig:EASFS2}
\end{center}   
\end{figure}

An typical example of an \EAS detected on the focal surface of a realistic \EUSO-like telescope is shown
in Figures~\ref{fig:EASFS} and~\ref{fig:EASFS2}.
   
%--------------------------------------------------------------------------------   
\section{Scientific and experimental requirements}   
\label{sec:SciObj}   
%--------------------------------------------------------------------------------

The following scientific objectives are used in this paper, as for a general-purpose experiment:
they provide the guidelines to derive the scientific requirements for the experiment.

The measured \UHECP energy spectrum has to be extended beyond $E \approx \sci{1}{20} \um{eV}$. 
A map of the arrival directions of \UHECP over the whole sky is required in
order to identify and localize compact sources, possibly studying their energy spectra. 
The mass composition of the primary particles has to be measured. 
The ultra-high energy neutrino flux has to be measured.

From these scientific objectives, a set of scientific requirements for the
experiment can be derived, see for instance~\cite{supereuso,petrosanta}.

%--------------------------------------------------------------------------------   
\subsection{Scientific requirements}   
\label{sec:SciReqs}   
%--------------------------------------------------------------------------------   
   
The following typical scientific requirements for \UHECP observation   
from space are used (see also~\cite{supereuso,petrosanta}), derived from the scientific
objectives.

Of course, different scientific objectives might be conceived, possibly leading
to different scientific requirements and different experiment design. The relations discussed in the rest of this paper may be adapted to
the specific situations.

All sky coverage is required. 

The angular resolution on the reconstructed primary \UHECP direction should be   
$\DD{\beta} \lesssim 3\degr $   
for a large enough sub-sample of events, in order to allow the identification of compact sources   
while taking into account the deviation induced by magnetic   
fields on charged particles.   
   
An energy resolution $\DD{E}/E \lesssim 0.3$ (including statistical and systematic contributions) is required.
In fact, the energy measurement errors distort the shape of a steeply falling
energy spectrum, because each energy bin collects more
mismeasured EAS from low energies than it does from higher energies. A resolution
smaller than $\DD{E}/E \lesssim 0.3$ does not
significantly modifies the energy-spectrum shape and does not smear the GZK feature~\cite{Albuquerque:2005nm}. 
   
The resolution on the depth of the \EAS maximum measurement has to be   
$\Delta{\Xmax} \lesssim 50 \um{g/cm^2}$ (accounting for the intrinsic variability of the \EAS,
depending on the \UHECP primary mass).   
  
An energy threshold \mbox{$ E_\mathrm{TH} \approx \sci{1}{19} \um{eV} $} is required,   
with a flat efficiency energy plateau at \mbox{$E \gtrsim  E_\mathrm{TH} $},    
in order to keep systematic effects well under control. 
In fact, near the low-energy threshold, the total \EAS detection efficiency
steeply changes. This  would lead to large  
systematic uncertainties due to the poor knowledge of the total \EAS detection
efficiency, which is determined from Monte-Carlo simulations only. 
A threshold at $E \approx \sci{1}{19} \um{eV}$ also ensures    
a fair overlap with the energy spectrum observed by ground-based experiments.   
Furthermore the telescope must be able to measure \EAS with energies up to    
$ E_\mathrm{MAX} \approx \scix{21} \um{eV} $, in order to study the \UHECP
energy spectrum in the trans-GZK region, by exploiting the larger
statistics with respect to ground experiments.

The instantaneous geometrical aperture, $ \GeoAperture $, is required to be one
order of magnitude larger than currently existing and/or planned ground-based
experiments, in order to increase by the same amount the number of collected
events:  
\mbox{$ \GeoAperture \gtrsim \scix{6} \um{km^2 \cdot sr}$}.

\subsection{Requirements for the experiment}   
   
%--------------------------------------------------------------------------------   
\subsubsection{Architecture of the telescope}   
%--------------------------------------------------------------------------------   
   
The required experimental telescope is made of a   
main digital camera, with cylindrical symmetry   
around the optical axis, operating in the near-UV. The digital camera must have a large aperture,   
a large \FoV and a fast and pixelized single photon detector; the latter
consists of   
the main optics, possibly a deployable optics, and the photo-detector on the
focal surface of the main optics. 
   
The main optics is made of the main mirror/lenses (large, lightweight and possibly segmented),   
corrector mirrors/lenses,   
the optical filters,   
the active control mechanism   
and the supporting structure.   
   
The photo-detector is located at the focal surface of the main optics which is
typically a strongly curved surface, due to the large \FoV of the optics.
It is made of    
a bi-dimensional array of photo-sensors,   
the light-collection system on the photo-sensor,   
the front-end electronics chip and ancillary electronics,   
the housing,   
the back-end, trigger and on-board data-handling electronics.   
   
Other ancillary systems, such as an Atmospheric Monitoring System and the    
Monitoring, Alignment and Self-Calibration system~\cite{EUSORedBook} are present.

%--------------------------------------------------------------------------------   
\subsubsection{Mission requirements}   
%--------------------------------------------------------------------------------   
   
The telescope has to observe the largest possible mass of atmosphere, with a large   
enough geometrical acceptance.   

Obstructions of the large \FoV, coming from the satellite structure, and parasitic lights coming from any other
source than the \EAS have to be minimized.    
   
The orbit should be accurately designed, in order to maximize the instantaneous
geometrical aperture, compatibly with
the required energy threshold (see Section~\ref{sec:Orbit}).   
   
The pointing accuracy is not a   
critical factor provided the absolute direction of the telescope axis is   
known for off-line use. The pointing direction must be known off-line to within    
a precision of $ \lesssim 0.5 \degr $, less than the   
expected angular resolution of the telescope.   
   
The expected rate of \EAS and background triggered events, affecting telemetry
and tele-commanding, will depend on the
final orbit and therefore a precise estimation is currently lacking. 
It will depend on the purity of the triggered sample, which is hardy estimable 
without preliminary measurements.   
   
The required lifetime is five years at least, in order to get a statistically
significant sample of events, at least one order of magnitude larger 
than expected from ground-based experiments.

%--------------------------------------------------------------------------------   
\subsubsection{Requirements for the telescope}   
%--------------------------------------------------------------------------------   

The telescope will operate in   
single-photon detection mode in the wavelength range  (WR)  
between $330 \um{nm}$ and $400 \um{nm} $,  
to detect the air scintillation light.   
Shorter wavelengths suffer absorption from the ozone layer. At higher wavelengths   
the scintillation emission is not significant and, moreover,
the larger the wavelength range the more difficult is to design an optics free
from chromatic aberrations. 
   
The faint signal requires    
high photon collection area and high photon detection efficiency, as well as   
low noise.   
Small cross-talk and after-pulse rates are required in order to obtain a good energy resolution.   

The dynamic range of the photo-detector should cover two orders of magnitude in signal intenisty, 
for an effective EAS detection between $ E_\mathrm{TH}$ and $ E_\mathrm{MAX}$.

The telescope should have an efficient and selective trigger system, to achieve a good background   
rejection on-board, and a powerful on-board data handling system as the
telemetry would be limited.   
   
The telescope has to be completely modular to reduce the risk of single point   
failures. Furthermore it should be robust against intense light sources like lightnings and man-made lights.   
   
The telescope has to satisfy the constraints relevant for a space mission including mass,   
power, volume, telemetry, as well as the many environmental factors.

%--------------------------------------------------------------------------------   
\subsubsection{Other technical requirements}   
%--------------------------------------------------------------------------------   

The interior of the telescope should be light-tight such that the   
parasitic lights impinging onto the photo-detector will be two orders of   
magnitude less than the expected night-glow background rate   
not to worsen the energy resolution.   
   
Other technical requirements, such as mechanical, thermal and electrical, are
also important, but they are not relevant to the current discussion.    
See~\cite{EUSORedBook} for details.

%--------------------------------------------------------------------------------   
\section{Definitions and assumptions}   
\label{sec:GenAss}   
%--------------------------------------------------------------------------------   
   
In this section the assumptions used in the paper are introduced.
The study is based on a typical hadronic \EAS viewed by a simple telescope,
looking downward the Earth, whose configuration and characteristics will be
defined in the rest of the paper.   
   
%--------------------------------------------------------------------------------   
\subsection{Some geometry definitions}   
%--------------------------------------------------------------------------------   
   
The \EAS properties depend, to a first approximation, only on the primary particle identity, its energy   
$E$ and its zenith angle, $\theta$.   

\begin{figure}[htb]   
\begin{center}   
\includegraphics[width=0.45\textwidth]{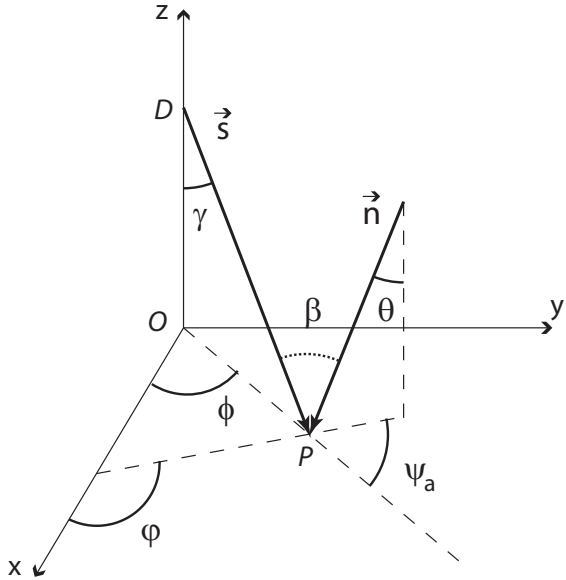}   
\end{center}   
	\caption{Definition of the angles with respect to a global coordinate frame $Oxyz$, 
                in the case of no tilt between the direction of the local nadir and the optical axis. 
                The $D$ is the position of the telescope, $O$ is the telescope nadir 
                and $P$ is the impact point of the \EAS on the Earth surface. 
                The \EAS direction is $\vec{n}$ (zenith angle $\theta$ and azimuth angle $\varphi$); 
                $\vec{s}$ the direction of the line-of-sight 
                (zenith angle $\gamma$ and azimuth angle $\FoVAzimuthAngle$); 
                $\beta$ is the angle between $\vec{n}$ and $\vec{s}$, while
                $\FoVShowerAzimuth = \varphi - \FoVAzimuthAngle$.   
        }   
\label{fig:FoVShowerDir}   
\end{figure}

The features of the \EAS image on the focal surface of the photo-detector
also depend on the location of the \EAS image inside the \FoV of the telescope,    
as defined by the angle between the optical axis and the line-of-sight, $\gamma$,   
and by the azimuth angle of the \EAS image projected at the Earth surface (see Figure~\ref{fig:FoVShowerDir}).   
The angle $\gamma$ will be called, in short, \emph{field-angle} in the rest of this paper, since it represents   
the visual angle inside the \FoV. Whenever the optical axis is coincident with
the local nadir, $\gamma$ is coincident with 
the zenith angle of the line-of-sight direction.   
The angle between the latter and the radial direction inside the \FoV is called
\FoVShowerAzimuth. If $\varphi$ is the \EAS direction azimuth angle, with
respect to a global reference system $Oxyz$, and \FoVAzimuthAngle\ is the azimuth angle of the line-of-sight   
in the \FoV, then $\FoVShowerAzimuth = \varphi - \FoVAzimuthAngle$.    

The unit vectors of the \EAS direction, $\hat{n}$, and of the line-of-sight, $\hat{s}$, in the frame $Oxyz$   
are respectively:

\begin{gather}   
	\hat{n}=    
	\{ - \sin \theta  \cos \varphi  ,\;- \sin \theta  \sin \varphi ,\;-\cos \theta\} \virgola    
	\nonumber \\    
	\hat{s}=    
	\{ \sin \gamma  \cos \FoVAzimuthAngle  ,\; \sin \gamma  \sin \FoVAzimuthAngle ,\;-\cos \gamma\}    
	\punto \nonumber   
\end{gather}   
   
If $\beta$ is the angle between the \EAS direction $\hat{n}$ and line-of-sight $\hat{s}$ from   
the telescope to the actual \EAS location one has:   
\begin{equation}   
	\cos(\beta) =    
	\cos (\gamma ) \cos (\theta )-\cos (\FoVShowerAzimuth) \sin (\gamma ) \sin(\theta )    
\punto   
\end{equation}

%--------------------------------------------------------------------------------   
\subsection{Reference conditions and general assumptions}   
\label{sec:Reference}   
%--------------------------------------------------------------------------------   
   
In this paper, unless otherwise specified, the conditions and parameters summarized in   
Table~\ref{tab:ReferenceConditions} will be used as a reference for most calculations.  
   
Several assumptions, discussed in detail in this section, are used.    
Some of these assumptions are somewhat optimistic with respect to the real   
conditions of the experiment.  

Is therefore emphasized once more that the results are minimal necessary, but possibly not sufficient,    
requirements for observation of \EAS from a space-based telescope.  

\begin{table*}[htb]   
\begin{center}   
%\begin{minipage}[h]{0.90\textwidth}   
\hfill{}   
\begin{tabular}{lcc}\hline   
Variable                            
	& Reference value     
	& Other values     
	\\    
	\hline   
	\hline   
Energy                              
	& $E=\un[10^{19}]{eV}$         
	&                  
	\\ 
Particle type
	& proton
	&
	\\   
Zenith angle                        
	& $\theta = 50\degr$     
	& $\theta = 30\degr,\:70\degr$                         
	\\   
Azimuth angle			    
	& $\FoVShowerAzimuth=90\degr$     
	& $\FoVShowerAzimuth=0\degr,\: 180\degr$    
	\\   
	\hline   
Telescope Orbital Height [km]   
	& $H=700$   
	& $H=400,\: 1000$   
	\\   
\FoV aperture (half-angle)                       
	& $\Gmax=20\degr$     
	& $\Gmax=15\degr,\: 25\degr$   
	\\   
Field-angle of observation   
	& $\gamma=15\degr$      
	& $\gamma=10\degr,\:20\degr$   
	\\   
Telescope Tilt Angle         
	& $\TiltAngle=0\degr$    
	&    
	\\   
Total photo-detection efficiency     
	& \PDE=0.1    
	&    
	\\   
	\hline   
Operating Wavelength Range (WR) [nm]	   
	& 330$\leq\lambda\leq$400   
	&    
	\\   
Random Night-glow Background in the WR [$\un{ph\cdot m^{-2}s^{-1} sr^{-1}}$]    
	& $B=\sci{5}{11}$    
	& $B=\sci{(0.3 \div 1.0)}{12}$    
	\\   
\hline        
\end{tabular}   
\hfill{}   
\caption{Parameters and conditions used as reference.}    
\label{tab:ReferenceConditions}   
%\end{minipage}   
\end{center}   
\end{table*}   
   
%--------------------------------------------------------------------------------   
\subsubsection{Satellite parameters}   
%--------------------------------------------------------------------------------   

The main parameters of the satellite    
are: its altitude above the Earth, $H$, (or, in more general terms, all the   
orbital parameters); the tilt angle $\TiltAngle$ between the   
optical axis and the local nadir (see Figure~\ref{fig:approach}).   
The orbital height $H$ is    
$ \un[400]{km} \lesssim H \lesssim \un[1000]{km}$ (see Section~\ref{sec:Orbit}).   

The telescope optical axis is parallel to the local nadir, that is $\TiltAngle=0$. 
Note that, if $\TiltAngle \neq 0$, the line-of-sight zenith angle is
different from the field-angle $\gamma$. 

%--------------------------------------------------------------------------------   
\subsubsection{EAS image properties}   
%--------------------------------------------------------------------------------   

Any \EAS, when observed from space, can be geometrically modeled as a point moving on a straight line   
at the speed of light. It therefore produces, on the curved focal surface of the
photo-detector, an image generated by a point moving on a line.

A reference energy \mbox{$E = \un[\sci{1}{19}]{eV}$} is used
(Table~\ref{tab:ReferenceConditions}), corresponding to the required energy
threshold.   
In fact the higher the energy of the \EAS, the better it can be reconstructed,
as more
photons are detected.

The \EAS direction and \FoVShowerAzimuth strongly affect    
the kinematics of the \EAS image and therefore its observability and its observed features.   
For several reasons, to be discussed later,    
only \EAS with a zenith angle    
$ 30\degr \lesssim \theta \lesssim 70\degr $ will be studied, with a reference
zenith angle $\theta = 50\degr $.
In fact a very inclined \EAS cannot be easily studied in a semi-analytical way, 
as it gives a very long image on the focal surface, such that the curvature
and the changes of the photo-detection characteristics of the
focal surface cannot be neglected.
On the other hand, an almost vertical \EAS produces a very short image   
on the focal surface and it is therefore hard to
reconstruct the \EAS.    
In order to describe an average behavior a reference angle $\FoVShowerAzimuth = \pm 90\degr$ will be used.   
   
Note that, when observing from space, up-going \EAS could be observed, with $\theta > 90\degr$.   
   
%--------------------------------------------------------------------------------      
\subsubsection{Atmospheric profiles}   
%--------------------------------------------------------------------------------   

The Linsley's parametrization of the US Standard Atmosphere~\cite{US-STD-76} is used.
This parametrization can be sometimes approximated with the simpler exponential density profile (isothermal atmosphere)   
as a function of the altitude, $h$, above the sea level~\cite{allen}:       

\begin{equation}   
\label{atmodensity}   
  \begin{split}   
    \rho(h) =    
    &\rho_0 \e^{-h/h_0} \\   
   \text{with}\quad h_0=\un[8.4]{km}, &\;    \rho_0=\un[1.2249]{kg/m^3}.      
\end{split}   
\end{equation}
   
%--------------------------------------------------------------------------------   
\subsubsection{Sphericity of the Earth}    
%--------------------------------------------------------------------------------   

A spherical Earth with radius   
\mbox{$R_{\oplus}= 6371 \um{km}$} is used.
The exact relation between the altitude above the   
Earth surface, $h$, and the distance $\ell$ measured along a straight   
line with zenith angle $\theta$ is:     

\begin{equation}\label{sphericalearth}   
\begin{split}   
\deriv{\ell}{h} &= \frac{h+R_{\oplus}}{\sqrt{h^2+2R_{\oplus}h+R_{\oplus}^2\cos^2\theta}}\\   
&\approx \left( 1- \frac{h}{R_{\oplus}}\tan^2\theta \right)\sec\theta    
\punto  
\end{split}   
\end{equation}

In the approximation of flat Earth,
when \mbox{$h \tan^2\theta \ll R_{\oplus} $}, the relation between $h$ and $\ell$ becomes:

\begin{equation}\label{flatearth}   
    \deriv{\ell}{h} = \sec\theta 
\quad .   
\end{equation}
   
It is also worth remembering the definition of the slant depth, $X$: 

\begin{equation}\label{eq:slantdepth}
        \deriv{X}{\ell}=\rho 
\virgola
\end{equation} 

plus the appropriate boundary condition on $X$ at some $\ell$.

%--------------------------------------------------------------------------------   
\subsubsection{EAS longitudinal and lateral profile}   
%--------------------------------------------------------------------------------   

The Gaisser-Hillas parametrization~\cite{gaisser-hillas,Pryke} of an hadron-induced \EAS describes    
the \EAS longitudinal profile as a gamma distribution:   

\begin{equation}\label{eq:gaisser-hillas}   
\begin{split}   
    N_\mathrm{ch}(X) & =    
	%\Nmax    
        %\pton{ \frac{ X - X_0 }{ \Xmax - X_0 } }^{ \pton{ \Xmax - X_0 } / \lambda }    
	%\exp\pqua{ \pton{ \Xmax - X } / \lambda }    
	%\\   
	%& =   
	\Nmax    
        \pton{ \frac{ X - X_0 }{ \Xris } }^{  {\Xris}/{\Lambda} }    
	e^{ \pton{ \Xris + X_0 - X }/{\Lambda} }    
	\\   
	&   
	\spazio\text{for $X \geq X_0$} \virgola   
\end{split}   
\end{equation}   

where $N_\mathrm{ch}(X)$, the number of charged particles at the slant depth $X$,    
is expressed in terms of:   
\Nmax, the number of charged particles at the \EAS maximum;  
\Xmax, the depth of the \EAS maximum; $X_0$, a fitting parameter; $ \Xris \equiv \Xmax - X_0 $ and  
\mbox{$\Lambda \simeq \un[65]{\um{g/cm^2}}$}, the interaction mean free path.   
This expression makes explicit the invariance   
of the \EAS longitudinal profile with respect to shifts of the parameter $X_0$.   
$X_0$ is often set at the first   
interaction point, although it is only a fitting parameter and it can also
take negative values. The $\Lambda$ parameter is rather energy   
independent and very similar for both proton and iron induced \EAS in   
the energy range under study.     
Finally $\Nmax \approx \kappa E$, with   
$\kappa \simeq 0.6/\mathrm{GeV}$~\cite{stanev}.   
   
Typical values for \Xmax and $X_0$ at \mbox{$E=\un[10^{19}]{eV}$} are given
in Table~\ref{tab:GaisseHillasParameters}, inferred from Monte-Carlo simulations
done with CONEX~\cite{Bergmann:2006yz}. 
   
\begin{table}[htb]   
	\centering   
	\begin{tabular}{c|c|c} \hline   
Parameter & Proton & Iron  \\ \hline \hline   
\Nmax     & \sci{6}{9} & \sci{6}{9}  \\   
$\Lambda$   & \un[65]{\um{g/cm^2}} & \un[65]{\um{g/cm^2}} \\   
\Xmax & $790\pm40$ g$/$cm$^2$ & $ 700\pm30 $ g$/$cm$^2$ \\   
$X_0$ & $35\pm5$ g$/$cm$^2$ & $10\pm 2$ g$/$cm$^2$ \\   
\Xris & $755\pm45$ g$/$cm$^2$ & $690\pm 42$ g$/$cm$^2$ \\ \hline   
	\end{tabular}   
	\caption{Typical Gaisser-Hillas parameters of a $E=10^{19}\un{eV}$ EAS for both proton and iron primaries.}   
	\label{tab:GaisseHillasParameters}   
\end{table}   
   
\begin{figure}[htb]   
    \centering   
        \includegraphics[width=0.48\textwidth]{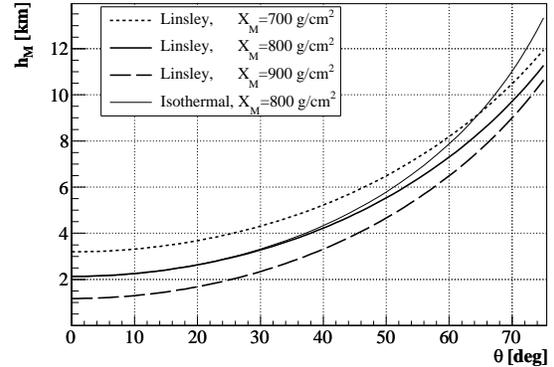}   
    \caption{The altitude of the \EAS maximum versus the \EAS zenith angle 
             for various atmospheric profiles.}   
    \label{fig:hmax}   
\end{figure}

A parameter strictly related to \Xmax is the altitude of the \EAS maximum, $h_\mathrm{M}$, which is fundamental   
for the reconstruction of \Xmax itself. It depends on \Xmax through the \EAS zenith angle $\theta$ and   
the atmospheric density profile.   
The values of $h_\mathrm{M}$, as a function of $\theta$,   
are shown in Figure~\ref{fig:hmax} for three different \Xmax using the   
Linsley's parametrization of the atmosphere density profile. The   
exponential (isothermal) profile for $\Xmax = 800 \um{g/cm^2}$ is also shown for comparison: it is clear that the   
choice of the atmosphere density profile does not strongly affect   
$h_\mathrm{M}$ up to $ \theta \approx 70\degr $.    
   
The lateral \EAS profile is not accounted for in this study, as it is hardly accessible when   
observing from space. 
Indeed, the typical \EAS lateral size is of about \mbox{one km}, corresponding
to the required spatial granularity (see Section~\ref{sec:Results}).    
The \EAS image is therefore only one or two pixels wide (for a perfect optics).   

%--------------------------------------------------------------------------------   
\subsubsection{Length of the \EAS track}    
%--------------------------------------------------------------------------------   

A typical hadronic \EAS of \mbox{$E \approx 10^{19} \um{eV}$}   
with \mbox{$\theta\lesssim 70\degr$} is seen on the focal surface of the
telescope    
as a few degrees long track, as discussed in detail in   
Section~\ref{sec:vislength}. This value is easily estimated by
computing the \EAS length between the first interaction point and the   
ground. The real length on the focal surface of the telescope is even shorter,
due to the predominance of background with respect to signal 
at the beginning of the \EAS.   
   
Within this short angular interval any possible change of the focal surface detection properties    
can therefore be neglected.     
Obviously this would not be true for   
a very inclined \EAS (\mbox{$\theta\gtrsim 70\degr$}), which may appear many degrees long and cannot be   
easily described in a simple parametrized form.   
   
%--------------------------------------------------------------------------------   
\subsubsection{Random Background}
%--------------------------------------------------------------------------------   

The background in the observation of \EAS from space 
is very complex and
produced by many sources (see Section~\ref{subsec:NoiseBackground} for more
details).

Only the random background due to the atmospheric night-glow and to the moon and
starlight has been taken into account, as it can be easily modeled.   
A uniform, isotropic and constant background,
on the space-time scale of the \EAS development is used.   
A reference value for the background of
\mbox{$ B \approx \sci{ 0.5^{+0.5}_{-0.2} }{12} \um{photons \cdot m^{-2} s^{-1} sr^{-1}} $}~\cite{baby},
in the wavelength range $\mathrm{WR}$, is used.   
The available measurements give, under certain circumstances,    
a value up to a factor two larger,    
depending on the conditions (including moon phases and cloudiness)~\cite{baby}.   
   
This random background, gives a large rate with respect to a typical
\EAS signal, and must be coped with (see Section~\ref{subsec:NoiseBackground}).
In this study it is assumed that the underlying background can be
subtracted away, in real time at the trigger level, in the region nearby the \EAS,   
by continuously measuring the average photon rate on a pixel-by-pixel basis.    
This is an optimistic but mandatory requirement to deal with the large random background.   
   
The expected large rate implies a small relative error on the   
background estimate (see Appendix~\ref{sec:NightGlowEst}).   
An appropriate statistical estimator to compare signal to   
background is therefore $ S / \sqrt{B} $.   

%--------------------------------------------------------------------------------   
\subsubsection{Air scintillation yield}    
%--------------------------------------------------------------------------------   

The \EAS develops in the atmosphere at an altitude $ h \lesssim   
\un[20]{km} $ (see Figure~\ref{fig:hmax}).  In this part of atmosphere the air scintillation yield in   
the wavelength range $\mathrm{WR}$ can be considered as nearly   
constant~\cite{kakimoto}, roughly  equal to $ Y \simeq (4.2 \pm 0.2) \um   
{photons}\cdot\um{particle^{-1}}\um{m^{-1}} $.    
Whenever a better precision is required the yield measurements
of~\cite{Nagano:2004am,bib:flash,Ave:2011ub} are used.    
   
%--------------------------------------------------------------------------------   
\subsubsection{Atmospheric transmission}    
%--------------------------------------------------------------------------------   

The atmospheric transmission is naively modeled taking into account Rayleigh scattering only.    
In fact Mie scattering is more important at low altitudes of a few km,    
while the \EAS predominantly develops above a few km   
of altitude (see Figure~\ref{fig:hmax}).    
Mie scattering is therefore ignored in this study at the current level of
approximation, as we are studying the minimal necessary requirements for observation.

The atmospheric transmission as a function of the zenith angle $\chi$, when taking into account Rayleigh
scattering only, can be roughly parametrized as:

\begin{gather}
	\label{eq:atmtra}  
       T_\mathrm{a}(\chi,h)  \approx
	\exp{\left[-\pton{\cfrac{\rho_0 h_0}{\Lambda_\mathrm{R}(\lambda)}}M(h,\chi)\exp{\pton{-\frac{h}{h_0}}}\right]}
\\
\text{with}\spazio
\cfrac{\rho_0 h_0}{\Lambda_\mathrm{R}} \approx 0.7 
\spazio\text{for}\spazio
\lambda = \un[337]{nm} \virgola \nonumber
\end{gather}

where: $h_\mathrm{0}$ and $\rho_0$ are, respectively, the atmosphere scale height
and the atmospheric density at the sea level (see equation~\eqref{atmodensity});
$\Lambda_\mathrm{R}(\lambda)$ is the Rayleigh mean free path at the wavelength
$\lambda$~\cite{Bucholtz:1995}; $M(h,\chi)$ is the air-mass function at the
zenith angle $\chi$~\cite{Young:1994}. 
When taking into account the Earth
curvature, the air-mass is given by the Chapman function
$\text{Ch}(h,\chi)$~\cite{bib:Grieder}.

Table~\ref{ta:AtmoTransm} shows the ratio, $\mathcal{R}$, between the   
atmospheric transmission for a zenith angle $\chi$ and   
the same quantity along the vertical direction for different altitudes.
%%%(without a tilt between the vertical   
%%%and the optical axis ($\TiltAngle=0$), $\chi=\gamma$).   
The table shows that at \mbox{$\chi \gtrsim 80\degr$} the atmospheric transmission 
from ground to infinity is reduced   
by more than one order of magnitude with respect to the vertical. This is   
crucial to evaluate of the effectiveness of a tilting of the   
telescope with respect to the nadir direction, as it will be discussed later
(Section~\ref{sec:tilting}).   
   
\begin{table}[htb]   
\begin{center}   
\begin{tabular}{r|c|c|c}                                              
\hline      
$\chi$ &  $ \mathcal{R}(h=0\um{km}) $ &  $ \mathcal{R}(h=5\um{km}) $ &  $ \mathcal{R} (h=10\um{km})$	\\   
\hline\hline   
   $  0\degr  $		& 	1.000	&  1.000	& 	1.000\\   
   $ 10\degr  $		& 	0.989	&  0.994	& 	0.997\\   
   $ 20\degr  $		& 	0.956	&  0.976	& 	0.986\\   
   $ 30\degr  $		& 	0.897	&  0.942        &  	0.968\\   
   $ 40\degr  $		& 	0.808	&  0.889   	&  	0.937\\
   $ 50\degr  $         & 	0.679	&  0.808	&	0.889\\
   $ 60\degr  $         & 	0.499	&  0.682	&	0.810\\
   $ 70\degr  $         & 	0.265	&  0.481	&       0.668\\
   $ 80\degr  $         & 	0.042	&  0.173	&	0.380\\
   $ 90\degr  $ 	&\sci{6.4}{-11} & \sci{2.4}{-6} & \sci{7.9}{-4}\\   
\hline   
\end{tabular}   
\end{center}   
        \caption{Ratio $ \mathcal{R}\equiv {T_\mathrm{a}(\chi,h)}/{T_\mathrm{a}(\chi=0,h)}$ between the   
                atmospheric transmission from the altitude $h$ above the ground to
                the telescope (at infinity), for a zenith angle $\chi$ and the same quantity for \mbox{$\chi=0\degr$}, 
                for three different altitudes. 
                All the quantities are evaluated for a wavelength $\lambda=\un[337]{nm}$ using equation~\eqref{eq:atmtra}.}   
\label{ta:AtmoTransm}   
\end{table}

The atmospheric geodesic refraction is usually negligible. In fact   
the correction for the refraction when observing an object   
at the Earth surface from a height $H$, at a zenith angle $\chi$ with   
respect to the nadir, is $\DD{\chi}\approx 62.37'' \tan\chi$, for   
$\chi\lesssim 80\degr$ at a wavelength $\lambda=\un[337]{nm}$~\cite{allen}.    
This implies $\DD{\chi}\approx 0.008\degr$ at $\chi=\Gmax=25\degr$.   
The effect of the refraction is therefore small with respect to the angular pixel size,
typically of $\sim 0.1\degr$ (see Section~\ref{sec:Results}).
However, the effect of atmospheric refraction might be important for a tilted   
telescope, when observing at large angles with respect to nadir: at $\chi=80\degr$ (maximum zenith angle    
with a tilt $\TiltAngle \approx 60\degr$), it is $\DD{\chi}\approx 0.1\degr$~\cite{allen}.
   
As we are studying the minimal necessary requirements for observation, we will
neglect multiple scattering and the effects of clouds, which are important in a real
experiment.
In fact these effects are difficult to study without a full Monte-Carlo
simulation and a precise experimental design.
Clear sky conditions are used. See~\cite{bib:esaf,Sokolsky:2003vj,AbuZayyad:2003af} for studies of the
effects of clouds.

%--------------------------------------------------------------------------------   
\section{Concept design and optimization of the experiment}   
\label{sec:Results}   
%--------------------------------------------------------------------------------   
   
In this section the requirements on the telescope, derived from the scientific
requirements, 
will be presented and discussed in detail.   
 
%--------------------------------------------------------------------------------   
\subsection{Orbit} \label{sec:Orbit}   
%--------------------------------------------------------------------------------   

The choice of the orbit of the telescope is a trade off between the energy threshold and the observed atmosphere volume.

The orbital height, $H$, (defined as the semi-major axis of an elliptical orbit)    
is one of the most important parameters.   
In fact a higher altitude provides a larger instantaneous geometrical aperture,    
but also a higher energy threshold   
because of the smaller \EAS signal received by the telescope.   
   
Practical and technical constraints may limit the orbital parameters.   
Indeed, orbits lower than about $ \un[300]{km} $ height suffer too much
atmospheric drag    
and require frequent re-boosts.   
On the other hand, orbits higher than about $ \un[1000]{km} $ height are
difficult to manage because the radiation   
environment undergoes a substantial change
due to the presence of   
Van Allen belts, whose high level of trapped radiation may adversely affect the
satellite operation.
   
An orbit with a variable $H$ is useful in order to extend the range of energy of the observed
\EAS: a satellite on an elliptic orbit spends more time at   
higher altitudes, gaining in the instantaneous geometrical aperture, but it also spends some   
time at lower altitudes, decreasing the energy threshold for \EAS observation, in an energy range where a long   
observation time might be not necessary, as the \UHECP flux at lower energy is larger.   
   
An orbit in the range of heights between
\mbox{$ H_\mathrm{MIN} \approx 400 \um{km}$} and \mbox{$H_\mathrm{MAX} \approx 1000 \um{km}$}
would scale up/down both the \UHECP energy range and the instantaneous
geometrical aperture by a factor:    
$ ( H_\mathrm{MAX} / H_\mathrm{MIN} )^2 \approx 6 $.    
On the other hand, linear distances, affecting the angular resolution and pixel
size,    
only scale as $H$, and not as $H^{2}$ (see Section~\ref{sec:AngResEst}).

The orbital height might be also varied by using different almost circular orbits during the mission   
lifetime (for example part of the time at a lower altitude and part of the time at a higher   
altitude).  Natural orbit decay might be exploited as well.   
   
\begin{figure}[htb]   
	\centering   
		\includegraphics[width=0.48\textwidth]{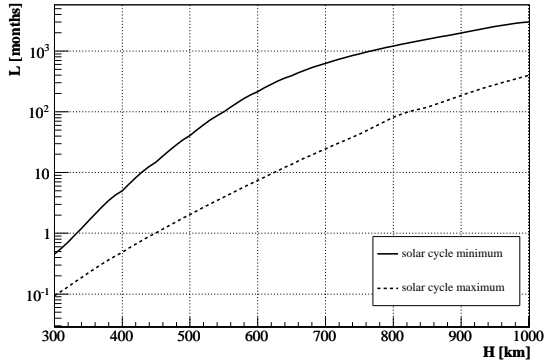}   
	\caption{Lifetime $L$ (in months) of a satellite (mass $\sim$ 3000 kg; cross-sectional   
		area $\sim$ 150 m$^2$) versus the orbital height $H$. Formulae and data for calculation can be found in~\cite{wertz}.}   
	\label{fig:lifetime}   
\end{figure}

The orbit lifetime    
depends on the orbital height and it strongly depends on the   
epoch of the Solar cycle (that is on the epoch of the Mission), which   
affects the atmospheric density (see   
Figure~\ref{fig:lifetime}).   
   
Another key parameter to keep in mind is the ballistic coefficient of the satellite,   
defined as $m/(C_\mathrm{d}A_\mathrm{sat})$, where $m$ is the satellite mass,
$A_\mathrm{sat}$ is its cross-sectional area    
and $C_\mathrm{d}\approx2.2$ is the atmospheric drag coefficient). It is expected to be rather low,   
due to the large expected area-to-mass ratio and therefore the atmospheric drag is expected to be large.   
   
Another important parameter is the    
\emph{observational duty cycle}, defined as the fraction of time with the telescope   
taking data. The orbit design must be further optimized to avoid as much   
as possible the light pollution from man-made sources at ground, from lightnings
(mainly present on the land in the equatorial regions), from auroras and other parasitic light sources.   
The effects of a short    
duty cycle can be recovered, in principle, by increasing the mission lifetime.   
   
The orbit can be also optimized to fly over specific fixed targets at ground, as often as   
possible. Such targets include ground-based \EAS experiments, weather   
stations and ground-based calibration sources.    
In particular, observations above ground-based \EAS experiments, namely the
Pierre Auger Observatory and Telescope Array, might be useful for cross-calibration.  

The coincidence rate of events detectable both by a space-based and a ground-based experiment,
strongly depends on the orbit. 
Order of magnitude estimates give the following results, taking a duty cycle of about 13\% for the space-based
experiment (an hybrid ground-based experiment has a duty cycle very close to 1).
In case of one passage per orbit above the ground target one has
about 5/15~events/year above 10/5~EeV, in coincidence with the PAO, and 
about 1/3~events/year above 10/5~EeV, in coincidence with the TA. 
In the case of a random passages above the ground target, these numbers go down
to 0.1/0.4~events/year for the PAO and 
0.03/0.1~events/year for the TA.

In fact, by designing a specific trigger only operational during the passages above the ground target, 
all the data collected by the telescope could be kept for offline analysis.
  
Clearly the design of the orbit is a crucial issue for such an experiment.
It is therefore preferred an experiment based on a free-flyer, whose orbit can
be optimized, to same extent, than an experiment, for instance, on the
International Space Station.
   
%--------------------------------------------------------------------------------   
\subsection{Optical system}   
%--------------------------------------------------------------------------------   
   
The two basic optical parameters, affecting both the performance and the
engineering of the experiment, are the Entrance Pupil (EP) area,
$A_\mathrm{EP}$, and the \FoV-angle $\Gmax$, defined as half the opening
angle of the circular cone generating the \FoV of the optics (see Figure~\ref{fig:approach}).    
   
The large required entrance pupil (a few meters diameter) calls for a ratio between the focal length and the
entrance pupil diameter as small as possible, in order to reduce the size of the
focal surface (see Section~\ref{sec:fsdimension}).    
   
%--------------------------------------------------------------------------------   
\subsubsection{Entrance pupil of the optics}   
%--------------------------------------------------------------------------------   

The entrance pupil area, $A_\mathrm{EP}$, or, equivalently, its diameter (also called \emph{optics aperture}), $D_\mathrm{EP}$,
is the basic parameter that affects   
the telescope sensitivity to low intensity signals.   
Increasing the optics aperture, to within the external and practical constraints,    
provides a guaranteed improvement of the telescope performance, since the
number of detected photons increases.    
   
The optics aperture also determines the size of the telescope and, consequently,   
the mass and the volume of the payload of the satellite. For large entrance
pupil diameters ($\gtrsim\un[3]{m}$),
a deployable optics   
is most probably required to fit within the space launch vehicle.

The total number of detected photons is roughly proportional to the solid angle subtended by the optics   
entrance pupil as seen by the \EAS   
(\mbox{$\Omega \approx A_\mathrm{EP}/H^2$}).   

The minimum value for $A_\mathrm{EP}$   
is basically set by the requirement on the energy threshold, while the    
upper limit is fixed by practical constraints such as money and technological
readiness. As the telescope is basically photon-limited, the entrance pupil must be as   
large as possible.   
   
In order to reach an energy threshold as low as 
\mbox{$E_\mathrm{TH}\approx 10^{19}\;\text{eV}$}, an entrance pupil diameter   
of the order of few meters at an orbital height between $400\um{km}$ and
$700\um{km}$ is required, (see Section~\ref{sec:TheoOptRequ}).   
   
%--------------------------------------------------------------------------------   
\subsubsection{\FoV of the optics}   
%--------------------------------------------------------------------------------   
   
In order to increase the instantaneous geometrical aperture
(Section~\ref{subsec:aperture}) and the number of detected EAS,   
either a higher orbit or a tilt of the telescope with respect to nadir might    
be more effective than increasing the \FoV. 
Indeed these two options might have two beneficial effects: keep the focal surface
size limited and ensure a better optics performance, as the performance of any
optics becomes typically worse when the \FoV increases.
   
The tight constraints of any space mission require the optical system to be
simple (few light-weight components). Such an optics has typically a relatively
large Point Spread Function (PSF), which varies with the position on the focal
surface. The size and shape of the PSF has a direct impact on the spatial
resolution at ground and the signal to background ratio. Therefore its optimization over
the focal surface is one of the critical aspects of the optical system
design. It is worthwhile to mention that a PSF much smaller than the required
resolution on ground would not be a realistic option, as the necessary number of pixels would become
too large. The optimal telescope design combines a sufficiently small PSF, and a pixel size of similar magnitude,
to limit the number of pixels.     
   
In the present discussion, a \FoV aperture (half-angle) $15\degr \le \Gmax \le
25\degr$, coming from the requirements    
on the instantaneous geometric aperture (see Section~\ref{subsec:aperture}), is used.
   
%--------------------------------------------------------------------------------   
\subsubsection{Reflective optics versus refractive optics}   
%--------------------------------------------------------------------------------   

One main drawback of a reflective system is that the center of curvature of the
focal surface is on the opposite side with respect to the incoming light.  The
filling of the focal surface with the array of photo-sensors is therefore more
difficult and the filling factor is worse than for a refractive system, with the
center of curvature of the focal surface is on the same side with respect to the
incoming light.  Furthermore a reflective system would suffer from the
obscuration due to the focal surface and the very limited room for the
photo-detector.  The latter might be a real issue, depending on the kind of
photo-sensors adopted.
   
On the other hand, a refractive system is possibly difficult to deploy    
and a complex supporting structure is required due to its large dimensions, causing some obscuration.
As for ground-based astronomical telescopes,
it is unpractical and difficult to build refractive telescopes above a certain size: an
entrance pupil diameter of a few meters is larger than the largest ground-based
refractive astronomical telescope.

On the basis of contemporary technology, a reflective system seems to be a more
viable option than a refractive one.
   
%--------------------------------------------------------------------------------   
\subsubsection{Size of the Focal Surface}\label{sec:fsdimension}   
%--------------------------------------------------------------------------------   
   
To a first approximation, the diameter of the focal surface, $D_\mathrm{PD}$, is a function of the   
\FoV-angle $\Gmax$ and of the focal length $f$:   

\begin{equation}   
        D_\mathrm{PD} \approx 2 f \sin \Gmax   
        \punto
\end{equation}   

Introducing the optics $f$-number:   

\begin{equation}   
 \Fnumb \equiv f/D_\mathrm{EP} \virgola   
\end{equation}   

where $D_\mathrm{EP}$ is the diameter of the entrance pupil, the above equation
can be re-written as:

\begin{equation}   
        D_\mathrm{PD} \approx 2 \Fnumb D_\mathrm{EP} \sin \Gmax   
        \punto
\end{equation}   

Note that for $\Gmax=30\degr$ and $\Fnumb = 1$ the focal surface is   
as large as the entrance pupil.    
   
A reasonable assumption for any real telescope is that the focal surface is   
not larger than the optics: if the latter is approximated by the optics   
entrance pupil, as it is desirable for a high-efficiency optics, this   
implies the rough estimate: \mbox{$2\Fnumb \sin \Gmax \lesssim 1$}.   
   
A detailed discussion of these and other topics about the optics can   
be found in~\cite{mazzinghi}.   
   
%--------------------------------------------------------------------------------   
\subsection{Effect of the observation angle}\label{sec:locationfs}   
%--------------------------------------------------------------------------------   
At the edge of \FoV ($\gamma \simeq \Gmax$) the triggering and    
reconstruction of the \EAS is more difficult than near the center of the \FoV.   
In fact, due to the larger \EAS distance, the flux of photons decreases as
$\approx \cos^{2}{\gamma}$ (the \emph{proximity factor}).

The effective entrance pupil typically
decreases as $\cos{\gamma}$, for a flat entrance pupil. It is the so-called
\emph{obliquity factor} of the optics: the entrance pupil area is seen reduced by a
factor $ \cos{\gamma} $ by incoming photons off-axis by an angle $\gamma$.    
   
Combining the obliquity and the proximity factors, the number of signal photons
falls as $\cos^3\gamma$ from pure geometrical effects.    

Under realistic assumptions, the optical system design is a trade-off among the
strict mass constraint, the complexity of the optics
and the large required aperture and \FoV. An optics with such features is likely to be
affected by anisoplanatism: the PSF is optimized for a certain field angle
$\gamma_\mathrm{best}$ (best focus, smallest size and almost round shape of the PSF, little glare)
but at field angles other than the optimal one, aberrations appear, the type and
strength of which depends on the optics implementation details.      
The choice of $\gamma_\mathrm{best}$ is driven by the scientific goals. If
$\gamma_\mathrm{best} = 0$, then the central region combines the best image
quality with a small obliquity factor, so that a lower energy threshold is achieved. 
On the other hand, if $\gamma_\mathrm{best} $ is close to the edge of the \FoV, it can be
used to partially compensate the worsening effects due the obliquity factor.    
   
In order to quantify the capability to focus incoming light onto a sufficiently small
region, it is useful to introduce the \emph{optical triggering efficacy} (with
dimensions of an area) as     

\begin{equation}\label{eq:trigefficacy}   
        \mathcal{E}^\prime_\mathrm{O}(\gamma) \equiv \frac{\Phi_\mathrm{b}(\gamma)}{I(\gamma)} \virgola   
\end{equation}      

where $I(\gamma)$ is the \emph{photon irradiance} of the signal (the number of
photons per unit time per unit area perpendicular to the line of
sight) and $\Phi_\mathrm{b}(\gamma)$ the number of photons incident on the focal
surface per unit time within a pixel bucket (that is a fiducial region around
the centroid of the PSF). The signal photons falling outside the pixel bucket
are usually too diluted to be associated to the \EAS signal, and are lost in the
background, contributing to the veiling glare,
especially at low energies. In the case of an ideal optics the
optical triggering efficacy becomes  

\begin{equation}\label{eq:idealtrigefficacy}
        \mathcal{E}^\prime_\mathrm{O} (\gamma) = A_\mathrm{EP}\cos\gamma \virgola
\end{equation}

where $\cos\gamma$ is nothing but the obliquity factor.
In real optical systems the optical triggering efficacy usually decreases faster than $\cos\gamma$,
for instance due to the absorption losses and off-axis aberrations.   
   
An \EAS detected at the edge of the \FoV would also suffer from a larger
atmospheric attenuation $T_\mathrm{a}$ (see equation~\eqref{eq:atmtra}), due to the
longer and more inclined path in the atmosphere.    
   
Generally speaking, the \FoV central region (more photons, better optical
quality but smaller geometrical
acceptance) is better for detecting low energy events with a low photon flux,
while the regions near the edge of the \FoV
(less photons, worse optical quality but larger geometrical acceptance) are
better for detecting high energy
events with a high photon flux.   
   
%--------------------------------------------------------------------------------   
\subsection{The Photo-Detector on the focal surface}   
%--------------------------------------------------------------------------------   

In this section rough basic estimates of an ideal photo-detector parameters are derived.
More precise estimates, for instance to quantify
the deviations from ideality, would require a full
Monte-Carlo simulation and a precise design of the main optics.

As discussed in section~\ref{sec:fsdimension}, a large focal surface area of a
few squared meters is required, to be filled with compact, robust and light-weight photo-sensors.

The pixel size is driven by two competing requirements: increasing $H$ (or
tilting the telescope) requires a smaller pixel size; increasing the optics
aperture (and therefore the focal length) implies a larger light spot on the
focal surface and therefore a larger pixel size.
   
The PSF of the optics has to match, approximately, the photo-detector pixel
size.  In fact, even though a finer pixel granularity would allow a better
reconstruction, provided that enough photons are collected, a trade-off with
cost and complexity is unavoidable.  On the other hand a pixel size much larger
than the PSF would not exploit all the quality of the optics (see
section~\ref{sec:pixelsize}).
   
The desired number of pixels, $N_\mathrm{pix}$, of the photo-detector can be   
estimated by the relation   

\begin{equation}\label{eq:nchans-e}   
    N_\mathrm{pix} \approx \cfrac{\pi H^2 \tan^2{\gamma}}{(\DD{L})^2 }   
    \virgola 
\end{equation}

where $\DD{L}$ is the linear granularity at ground.
   
The required photo-detector pixel size, $d$, corresponding to observing   
a length $\DD{L}$ on the Earth surface, is then   

\begin{equation}\label{eq:pixels-1}   
    d \approx \cfrac{f \DD{L}}{H}   
    \punto   
\end{equation}   
   
The photo-detector surface has to approximate the focal surface of the
optics, which is expected to be strongly curved due to the large \FoV and the
impossibility to build an optimized optics with many components.
The focal surface, to a first approximation, can be assumed to have a spherical
shape (the exact shape for a spherical primary mirror), with radius equal to
$f$ and maximum angular aperture \Gmax.   
Its area, $A_\mathrm{PD}$, is given by   

\begin{equation}\label{eq:adet}   
    A_\mathrm{PD} = 2 \pi f^2 \pton{1-\cos{\Gmax}}   
    \punto   
\end{equation}

The approximate maximum number of pixels which can be fitted on
the focal surface of the photo-detector is   

\begin{equation}\label{eq:nchans-fs}   
    N_\mathrm{pix} \approx \cfrac{ A_\mathrm{PD} }{d^2}   
    \punto   
\end{equation}   
    
Alternatively the desired pixel size on the focal surface
can be estimated, in terms of the focal surface parameters and the desired
number of pixels, by the relation   

\begin{equation}\label{eq:pixels-2}   
    d \approx \sqrt{\cfrac{ A_\mathrm{PD}}{N_\mathrm{pix}}}   
    \punto   
\end{equation}   
   
Note that, to the present level of approximation and with the present
parameters, the two equations~\eqref{eq:pixels-1} and ~\eqref{eq:pixels-2} are roughly equivalent,   
given the equations~\eqref{eq:nchans-e} and~\eqref{eq:adet}.   

The angular granularity of the photo-detector, $\DD{\alpha}$, is given by the
relations   

\begin{equation}\label{eq:angulargran}   
        \Delta\alpha \approx \cfrac{\DD{L}}{H} \simeq \cfrac{d}{f}    
\punto   
\end{equation}   

The solid angle coverage of every pixel, $\Delta\Omega$, is given by   

\begin{equation}   
        \Delta\Omega \approx \cfrac{(\DD{L})^2}{H^2} \simeq \cfrac{d^2}{f^2}    \simeq
        \pqua{\Delta\alpha}^2    
\punto   
\end{equation}   

The optics also determines the distribution of incidence angles of the photons
on the focal surface, which has some impact on the photo-detector design, because
the detection efficiency of the photo-sensor in general depends on it.    
The
marginal ray angle is determined by the \Fnumb of the optics and is given by
the relation   

\begin{equation}\label{eq:marginalAngle}   
        \tan{\gamma_\mathrm{marg}} \approx \cfrac{1}{2 \Fnumb}    \virgola   
\end{equation}   

which leads to a large spread of incidence angles of photons on the focal surface.

In order to reduce the effect of defocusing on the large and curved focal
surface a good fit between the ideal focal surface of the optics and the real
shape of the photo-detector must be implemented. In general it is not trivial at
all to obtain a good fit, because the optimized focal surface of the optics has
a complex geometrical shape while the photo-sensor modules typically have a flat
photo-sensitive surface with a simple geometrical shape.  
The approximate maximum
defocusing in the direction parallel to the focal surface, $\Delta w$, produced
by a small displacement $\Delta z$ in the direction perpendicular to the focal
surface, is given by the relation

\begin{equation}\label{eq:defocusing}   
    \Delta w \approx \Delta z \tan{\gamma_\mathrm{marg}} = \cfrac{\Delta z}{2 \Fnumb} \punto   
\end{equation}   

It should be $\Delta w \ll d $, which sets a requirement on the maximum value of
$\abs{ \Delta z}$.

%--------------------------------------------------------------------------------
\subsection{The detected air scintillation signal}
\label{sec:detectedsignal}
%--------------------------------------------------------------------------------

It is worthwhile to recall the relation between the signal from the \EAS detected by the
telescope and the most basic parameters of the \EAS, of the atmosphere and
of the telescope itself. All
the formulas presented in this section depend on the photon
wavelength, but the explicit dependence will be omitted.

An \EAS develops at the speed of light $c$, starting from the first interaction point $\ve{x}_0$,
along the longitudinal coordinate $\ell = c(t-t_0) = |\ve{x}(t)-\ve{x}_0|$,
where $\ve{x}(t) $ or $\vec{x}(\ell(t)) $ is the position as a function of time.
Let $\ve{x_\mathrm{T}}$ be the position of the telescope, assumed to be constant during the 
EAS development.

Three main factors contribute to the number of signal counts, $\mathcal{N}$: the air
scintillation signal $N_\mathrm{sc}$, the atmospheric transmission from the \EAS
to the telescope  
$T_\mathrm{a}$ and the telescope photo-detection efficacy
$\mathcal{E}_\mathrm{tel}$, with the dimensions of an area.
Let $\Delta \ell$ be a short segment of the \EAS located at $\ve{x}(\ell)$. 
The characteristics of the three previous factors can be taken as constant over $\Delta\ell$ and therefore:

\begin{equation}
      \mathcal{N} = \frac{\mathcal{E}_\mathrm{tel}  T_\mathrm{a} N_\mathrm{sc}}{4 \pi|\ve{x}-\ve{x_\mathrm{T}}|^2 } \punto
\end{equation}

The number of air scintillation photons generated by the \EAS segment $\Delta\ell$ is given by

\begin{equation}
	N_\mathrm{sc} = N_\mathrm{ch}(\ve{x})    Y(\ve{x}) \Delta \ell \virgola
\end{equation}

where $N_\mathrm{ch}(\ve{x}) $ is the number of charged particles in the \EAS and $Y(\ve{x})$ is the air scintillation yield. 
In principle, $Y(\ve{x}) $ also depends on the energy of the charged particles,
which varies with the energy and zenith angle of the \EAS.
However, at this level of approximation, only the dependence on the spatial
position will be taken into account.

The atmospheric transmission, 
$T_\mathrm{a}$, depends on the emission point \ve{x} and the telescope position
$\ve{x}_T$, or equivalently on the angle $\chi$, and the altitude of the
emission point $h(\ell)$, see equation~\ref{eq:atmtra}.

The overall \emph{photo-detection efficacy} of the telescope,
$\mathcal{E}_\mathrm{tel}$, factors into the optical triggering efficacy,
$\mathcal{E}'_{O}$, the optical filters efficiency, $\varepsilon_\mathrm{f}$,
and the overall photo-detection efficiency, $\PDE$: 

\begin{equation}\label{eq:telefficacy}
	\mathcal{E}_\mathrm{tel}(\gamma) = 
        \mathcal{E}'_{O}  \varepsilon_\mathrm{f}  \PDE \punto
\end{equation}

The overall photo-detection efficiency $\PDE(\gamma)$ is
the probability that a photon reaching the photo-detector will fire the
front-end electronics producing a recorded photon hit. It is one of the most
important parameters affecting the performance and depends itself on many
parameters, in particular on the field-angle $\gamma$.  For instance, the
incidence angle of the incoming photons at the photo-detector entrance window
changes with the field-angle. Typically, the larger the incidence angle, the
lower the photo-detector efficiency is. The smaller the \FoV of the optics the
larger the incidence angles are, according to relation~\eqref{eq:marginalAngle}.

For space-qualified standard PMT, already tested in the space environment,
$\PDE=0.12\div 0.15$. Even in the most optimistic scenario it is not realistic to expect more
than a factor three improvement of the $\PDE$ in the
near future, provided that successful efforts can be devoted to the development of suitable
photo-sensors with higher quantum efficiency, that is nowadays the main source
of inefficiency.  However also the geometrical acceptance and filling factor of
the array of photo-sensors are crucial issues.   

Recently new photocatodes for photo-multiplier tubes have been developed. They have a quantum efficiency about
twice the one of traditional bi-alkali photo-cathodes, see for
instance~\cite{Nakamura:2010zzk}.
Moreover many different concepts of solid state photo-sensor devices are being
actively developed by many research groups. These devices, which come with
different names (GAPD, SiPM, ...), aim to reach a total photo-detection
efficiency of $\PDE=0.4 \div 0.6$ (see, for instance,~\cite{ProcNDIP} for the
most recent developments).
However, in order to be ready for \EAS observation from space, some R/D is still required to improve on such parameters as
dark count rate, filling factor and quantum efficiency in the near-UV.

Other effects may affect the overall photo-detection efficiency, with
efficiencies typically very close to one. 
These effects include, for instance, reflection and absorption at the various optical elements,
the filling factor of the focal surface with the real photo-sensor modules,
threshold and pile-up effects in the front-end electronics.
However the product of many
efficiencies close to one can decrease the overall efficiency by a substantial
amount. However, as they are already very
close to one, there is little hope to gain anything substantial there.

The number of signal photons produced by a segment $\Delta \ell$ of the \EAS, at
the coordinate \ve{x},  
and detected by the telescope is thus: 

\begin{equation}\label{eq:Signal}
  \mathcal{N} = 
       \cfrac{ N_\mathrm{ch} Y \DD{\ell} }{4 \pi|\ve{x}-\ve{x_\mathrm{T}}|^2 } 
	T_\mathrm{a}(\chi,h)
	\mathcal{E}^\prime_\mathrm{O}(\gamma)
	\varepsilon_{\mathrm{f}}
	\PDE
\punto
\end{equation}

This equation can be used after defining the fiducial region on the focal
surface such that photons are considered to be 
detected photons, that is photons that are close enough to the centroid of the
image spot, see Section~\ref{sec:locationfs}. 
The photons outside the fiducial region just contribute to the veiling glare, reducing the image contrast.

It is worthwhile to note that for faint signals (a few photons per pixel) the errors in the equation~\eqref{eq:Signal}
will be typically dominated by the Poisson statistical fluctuations. 
In fact, for \EAS with a number of detected signal photons $\mathcal{N}\approx 100 $, less than 10 photons per pixel are expected with a relative statistical error of the order of $0.3$, larger than the expected systematic errors of about 15\% (see Appendix~\ref{sec:AngResEst}).

%--------------------------------------------------------------------------------
\subsubsection{Multi-photon detection factors}
%--------------------------------------------------------------------------------

Other more complex factors, 
involving multi-photon correlations on the same pixels and correlations between
photons in neighboring pixels,
will affect the \EAS detection efficiency.
These include cross-talk, the front-end/read-out electronics and trigger efficiency
and the event reconstruction efficiency.
These effects are very difficult to estimate without a full Monte-Carlo simulation~\cite{bib:esaf,TheaThesis,PesceThesis}, 
as they depend on the contribution of all the photons at once.

%--------------------------------------------------------------------------------
\subsubsection{Signal and background roll-off with field-angle}
%--------------------------------------------------------------------------------

Let us assume an ideal optics: $ \mathcal{E}^\prime_\mathrm{O}(\gamma) = A_\mathrm{EP}\cos{\gamma}$.

The field-angle dependence of the signal, $S(\gamma)$, for a given source is given by 

\beq 
\begin{split}
	S(\gamma) &\propto
	T_\mathrm{a}(\gamma,0) \mathcal{E}^\prime_\mathrm{O}(\gamma) \cos^2{\gamma} \\
	&\longrightarrow \gamma (0\degr \rightarrow 25\degr) : (1.0 \rightarrow 0.69)
\punto
\end{split}
\eeq

The field-angle dependence of the random background, $B(\gamma)$, is 

\beq
\begin{split}	
        B(\gamma) &\propto  
	\mathcal{E}^\prime_\mathrm{O}(\gamma) \\
	&\longrightarrow \gamma (0\degr \rightarrow 25\degr) : (1.0 \rightarrow 0.90)
\punto
\end{split}
\eeq

Both signal and random background decrease at increasing field-angles, but in a
different way: $S$ decreases faster than $B$ 
with $\gamma$. In fact 
the field-angle dependence of the ratio between the signal and the square root
of the random background is:

\beq
\begin{split}
	\cfrac{ S }{ \sqrt{B} } (\gamma)&
        \propto
        T_\mathrm{a}(\gamma,0)\sqrt{\mathcal{E}^\prime_\mathrm{O}(\gamma)}\cos^2{\gamma}  
        \\
	&
        \longrightarrow  \gamma (0\degr \rightarrow 25\degr) : (1.0 \rightarrow 0.73)
\virgola
\end{split}
\eeq

The dependence on the field-angle of both the signal and
$S/\sqrt{B}$ makes the increase of the optics \FoV useless above a
certain extent for a nadir pointing telescope. 
Similar considerations apply to tilted instruments (see section~\ref{sec:tilting}).  

The trigger settings require tuning as a function of the field-angle, 
that is as a function of the radial distance on the focal surface. 

One important result is that a detection energy threshold about two times higher is expected 
for events detected at \mbox{$\gamma = 25\degr$} with respect to events detected 
on-axis, because 
the efficiency curve as a function of the energy roughly scales
as the inverse number of signal photons detected~\cite{bib:esaf,TheaThesis,PesceThesis}.

%--------------------------------------------------------------------------------
\subsection{Number of detected photons and energy resolution}
\label{sec:NumPh}
%--------------------------------------------------------------------------------

The desired energy resolution (\mbox{$\Delta E/E \sim 0.2 \div 0.3$}) calls for
a relative error due to the Poisson statistics of 
the number of detected photons, $\mathcal{N}$, not larger than $ 1/\sqrt{\mathcal{N}} \sim\pton{0.10 \div  0.15} $, 
assuming an equal contribution from Poisson statistical and other
systematic errors. This
implies that at least $\mathcal{N} \simeq 100$ photons
must be detected from any \EAS at the lowest energies, with full detection efficiency.
Obviously, the number of detected photons impacts significantly on all the other
observables too.

\begin{figure}[htbp]
	\centering
		\includegraphics[width=0.49\textwidth]{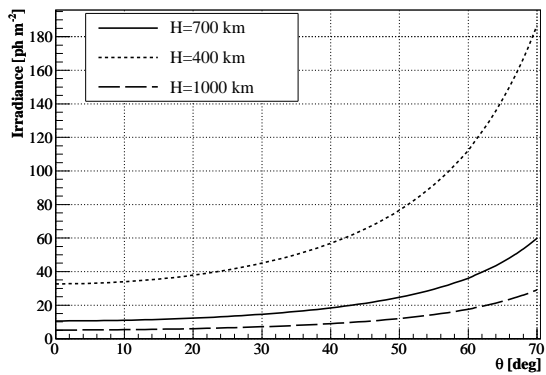}
		\caption{Signal time-integrated irradiance in the reference
                  conditions of Table~\ref{tab:ReferenceConditions}, as a function of $\theta$. }
		\label{fig:irrvstheta3h}
\end{figure}

The time-integrated irradiance of the signal reaching the telescope, as a
function of the zenith angle $\theta$, is shown 
in the Figure~\ref{fig:irrvstheta3h}.
For an ideal optics and the reference conditions of Table~\ref{tab:ReferenceConditions},
the required overall triggering efficacy
and the minimum telescope diameter can be inferred from the time-integrated
irradiance (see section~\ref{sec:Estimates}). 
This result does not depend on the implementation of the optical system and the
photo-sensor.
The required overall triggering efficacy for observing the same \EAS scales as $H^2$.

It follows that a minimum entrance pupil diameter
$D_\mathrm{EP}\approx\un[8]{m}$ is required to observe $\mathcal{N}\simeq 100$
signal photons, in the reference conditions of Table~\ref{tab:ReferenceConditions},
at the energy threshold.

%--------------------------------------------------------------------------------   
\subsection{The length and duration of the visible \EAS image}\label{sec:vislength}   
%--------------------------------------------------------------------------------   
   
The angle subtended by the visible \EAS image on the focal surface of the   
telescope and its time duration, which two important quantities, necessary for the
  experiment design,   
are easily estimated by determining the first and   
last detected points of the \EAS image.   
   
If one assumes to be able to subtract all the background and neglects the contribution of
the Cherenkov light reflected by the Earth surface, the shape of   
the detected photon hits is well described by the Gaisser-Hillas   
function (equation~\eqref{eq:gaisser-hillas}), which is a Gamma distribution in the variable   
\mbox{$ \widehat{X} \equiv X - X_0 $}:   

\begin{gather}\label{eq:gamma}   
	g(\widehat{X}) =    
	\frac{b^{-a}{\widehat{X}}^{a-1}}{\Gamma(a)} e^{{ -\frac{\widehat{X}}{b}   }} \propto   
        {\widehat{X}}^{{ \frac{\Xris}{ \Lambda} } }   
	e^{{  - \frac{ \widehat{X}  }{ \Lambda } }}   
	\\   
	\text{where} \quad a \equiv \frac{\Xris}{\Lambda} + 1,   
	\quad   
	b \equiv \Lambda,   
	\quad   
	\Xris \equiv \Xmax - X_0   
\punto \nonumber   
\end{gather}   

The mean and the   
standard deviation of this gamma distribution are:

\begin{gather}   
	\left\langle \widehat{X} \right\rangle = a b = \Xmax - X_0 + \Lambda
        \virgola
        \nonumber 
        \\   
	\sigma_{\widehat{X}} = b \sqrt{a} = \sqrt{\Lambda(\Xmax - X_0 + \Lambda)}
        \punto
        \nonumber   
\end{gather}   
   
\begin{figure}[htbp]
   \centering       \includegraphics[width=0.48\textwidth]{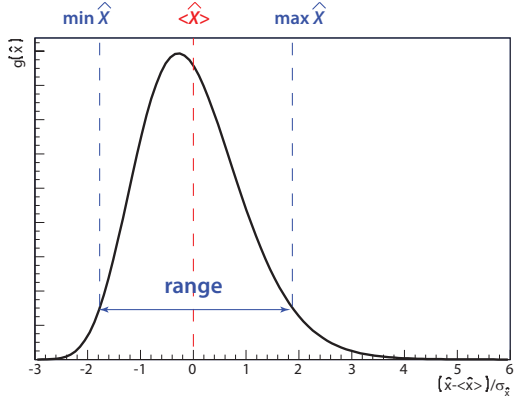}
       \caption{An example of Gamma function $g(\hat{X})$, with the indication of the minimum and maximum
values and the range of $\hat{X}$.}
   \label{fig:GammaBase}
\end{figure}

The number of detected photons, $\mathcal{N}$, can be estimated starting from
the parameters of the \EAS and telescope. Moreover they follow
the probability distribution of equation~\eqref{eq:gamma}.
The minimum and maximum value and the range of $ \widehat{X}$ (see Figure~\ref{fig:GammaBase}) as a function of
$\mathcal{N}$ can be easily determined from simulations of Gamma distributions,   
using the physical parameters given in section~\ref{sec:Reference}, 
Table~\ref{tab:GaisseHillasParameters}. Simulations give good enough results 
for the present study; they are easier and faster than a precise analytical
study (see for instance~\cite{KS}).
A proton primary particle was used,
but it has been checked that the results do not differ significantly for other primary hadrons.
For instance, in case of an iron \EAS at \mbox{$E=10^{19}$ eV}, the
range for \mbox{$\mathcal{N}=100$} is $4.9\sigma$ instead of $5.0\sigma$.

\begin{figure}[htbp]
	\centering
		\includegraphics[width=0.48\textwidth]{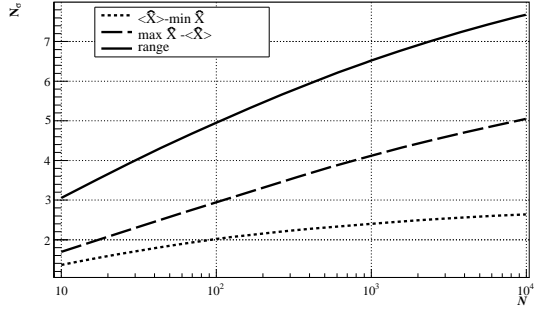}
		\caption{The minimum and maximum values and the range of 
                  $\widehat{X}$ as a function of $\mathcal{N}$ for a Gamma
                  distribution with parameters corresponding to the reference
                  \EAS (Table~\ref{tab:ReferenceConditions}).} 
		\label{fig:GammaStats}
\end{figure}

The minimum and maximum value and the range of $ \widehat{X}$ as a function of $\mathcal{N}$   
are given in Figure~\ref{fig:GammaStats} and Table~\ref{tab:GammaStats}. 
As it is required to observe \mbox{$\mathcal{N}\simeq 100$} photons at the energy
threshold, the range turns out to be about $5\sigma$.    

As expected, the results show that, due to the Gamma-like shape of the
Gaisser-Hillas function, the range of the detected photon
distribution (that is the observed image length) does not change by a   
large amount as a function of the number of detected photons.   
In fact, as Figure~\ref{fig:GammaStats} shows, the range is roughly a logarithmic function
of the number of detected photons.
   
\begin{table}[htb]   
    \centering   
        \begin{tabular}{cccccccc}   
	\hline	   
	$\mathcal{N}$	&	& range		   
	& $\left\langle \widehat{X} \right\rangle$-min $\widehat{X}$ & & &   
	& max $\widehat{X}$-$\left\langle \widehat{X} \right\rangle$ \\   
	\hline	\hline   
	$10$	&	& $(3.2 \pm 0.1)\sigma$	 & $1.4\sigma$ & & & & $1.8\sigma$	\\   
	$100$	&	& $(5.0 \pm 0.1)\sigma$	 & $2.0\sigma$ & & & & $3.0\sigma$	\\   
	$1000$	&	& $(6.5 \pm 0.1)\sigma$	 & $2.4\sigma$ & & & & $4.1\sigma$ 	\\   
	$10000$	&	& $(7.9 \pm 0.1)\sigma$  & $2.6\sigma$ & & & & $5.3\sigma$ 	\\   
	\hline   
        \end{tabular}   
        \caption{Estimation of the range of a Gamma distribution in the reference conditions of Table~\ref{tab:ReferenceConditions}.}   
\label{tab:GammaStats}   
\end{table}   
   
\begin{figure*}[htb]   
	\centering   
\begin{tabular}{cc}   
	\includegraphics[width=0.48\textwidth]{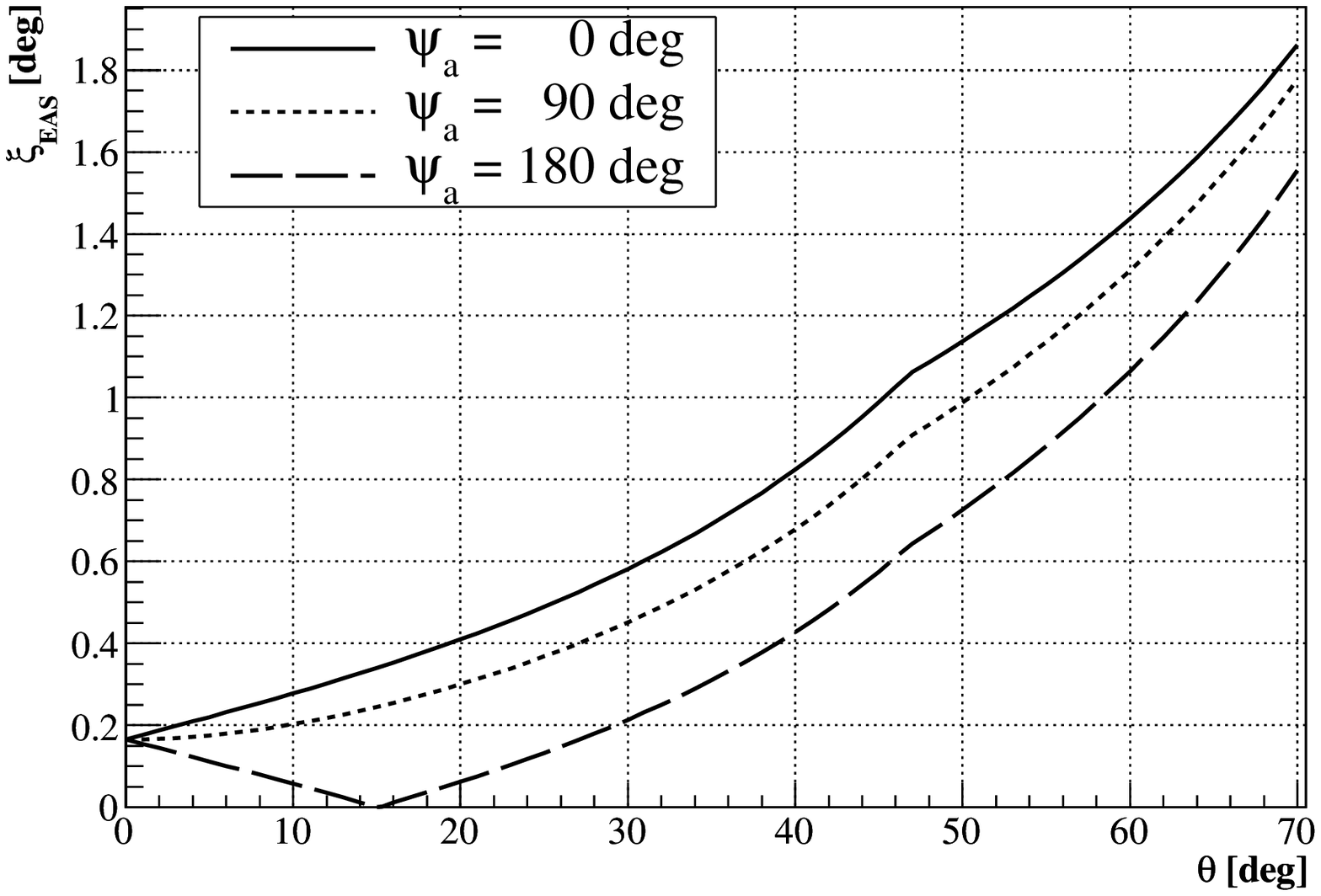} &   
	\includegraphics[width=0.48\textwidth]{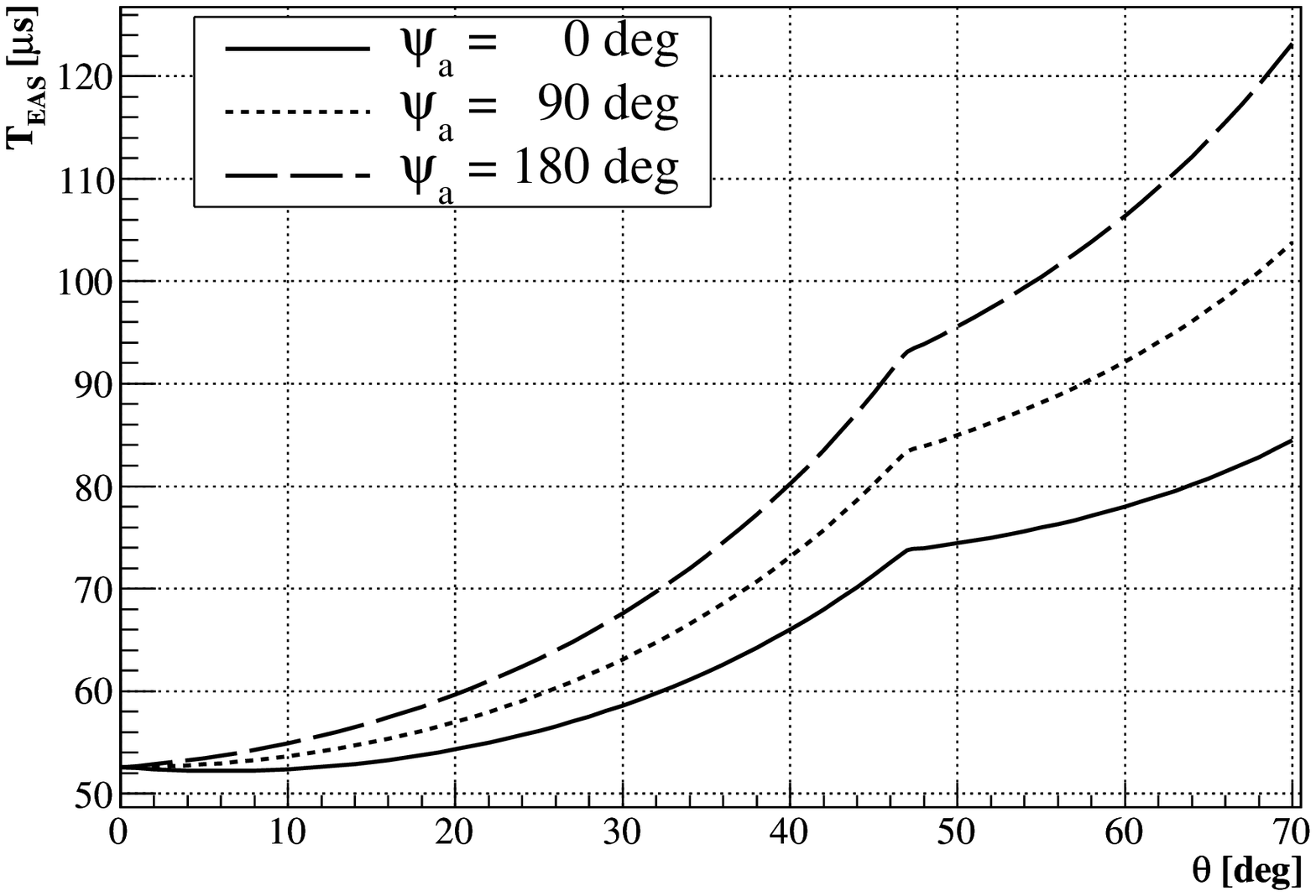} \\   
	(a)  &  (b) \\   
\end{tabular}   
		\caption{Angular extension (a) and time duration (b) of a $E=10^{19}$ eV proton \EAS image as a function of $\theta$ ($H=700\;\text{km}$ and $\gamma=15\degr$). }\label{fig:angle3psi}   
\end{figure*}

\begin{figure*}[htb]   
	\centering   
\begin{tabular}{cc}   
		\includegraphics[width=0.48\textwidth]{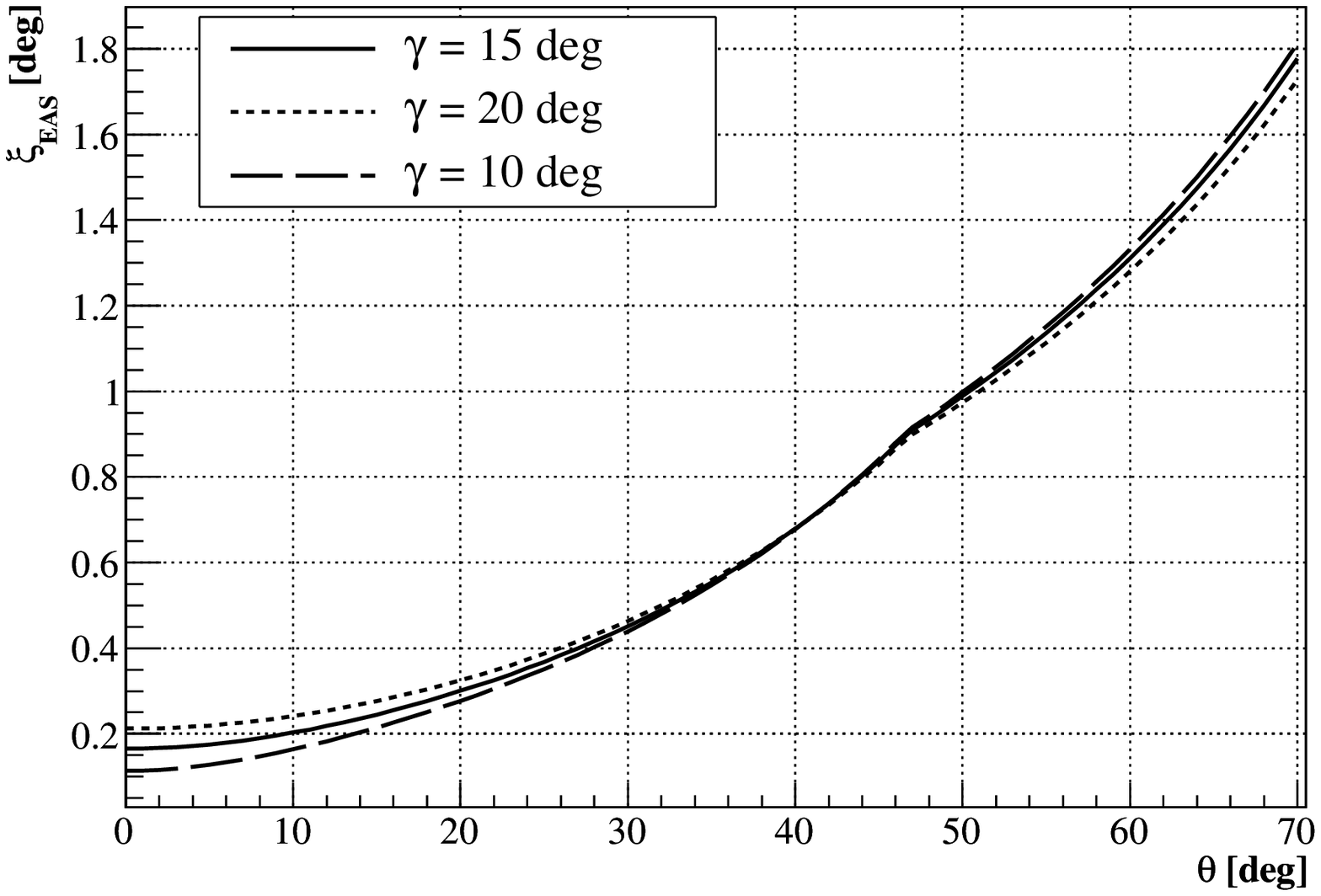} &   
\includegraphics[width=0.48\textwidth]{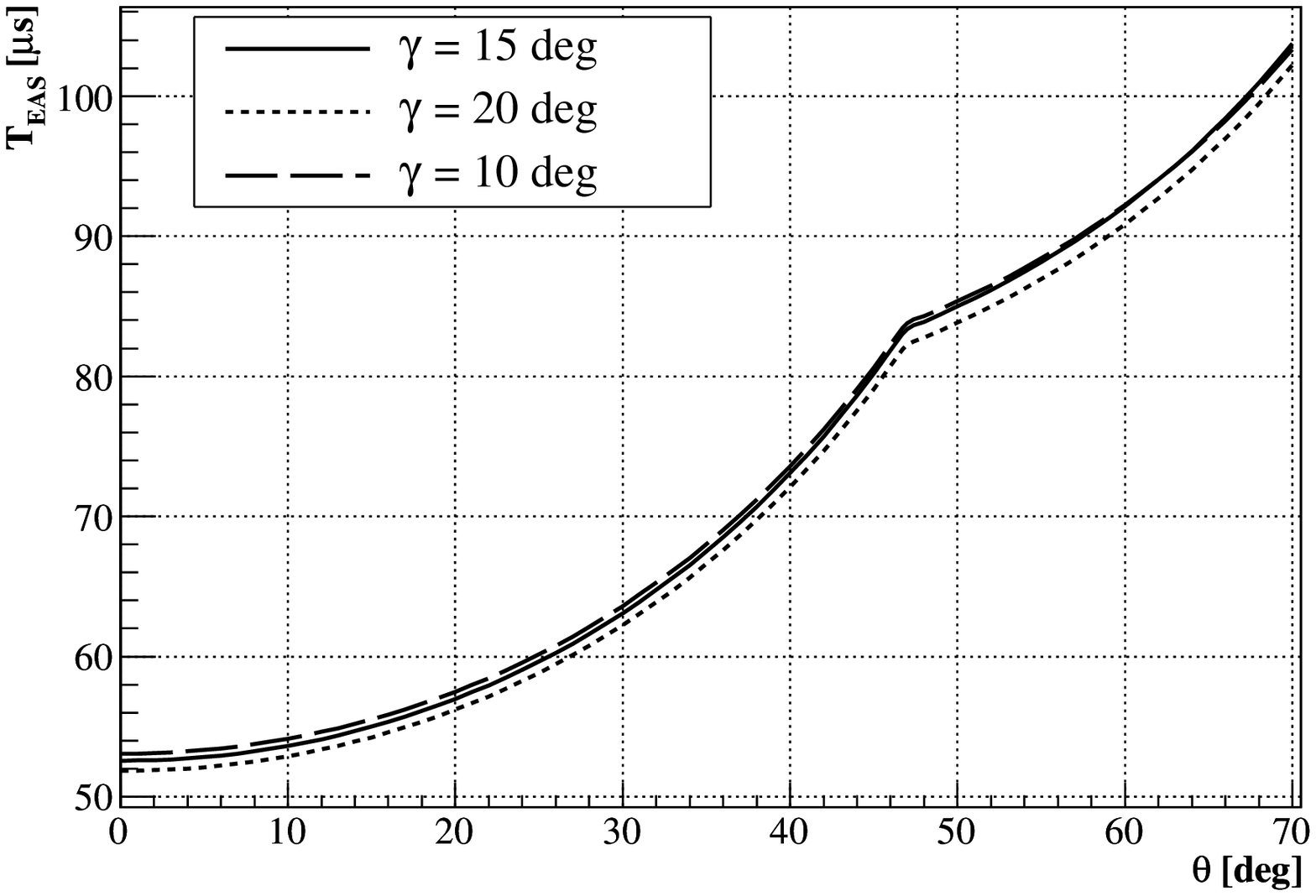} \\   
	(a)  &  (b) \\   
\end{tabular}   
		\caption{Angular extension of a $E=10^{19}$ eV proton \EAS image on the focal surface as a function of $\theta$ ($H=700\;\text{km}$    
and $\psi_\mathrm{a}=90\degr$).  }   
\label{fig:anglevstheta3gamma}   
\end{figure*}

From the above results, it is easy to estimate the angle $\EasAngle$ subtended by the \EAS   
image and the \EAS image duration $\EasTimeLength$.    
The dependence on the \EAS zenith angle $\theta$ of the angular extension 
and time duration of a $E=10^{19}$ eV proton \EAS image
are shown respectively in   
Figures~\ref{fig:angle3psi}$\div$\ref{fig:anglevstheta3gamma} for different
heights of the satellite and different values of $\FoVShowerAzimuth$.

\begin{figure*}[htb]   
	\centering   
\begin{tabular}{cc}   
		\includegraphics[width=0.48\textwidth]{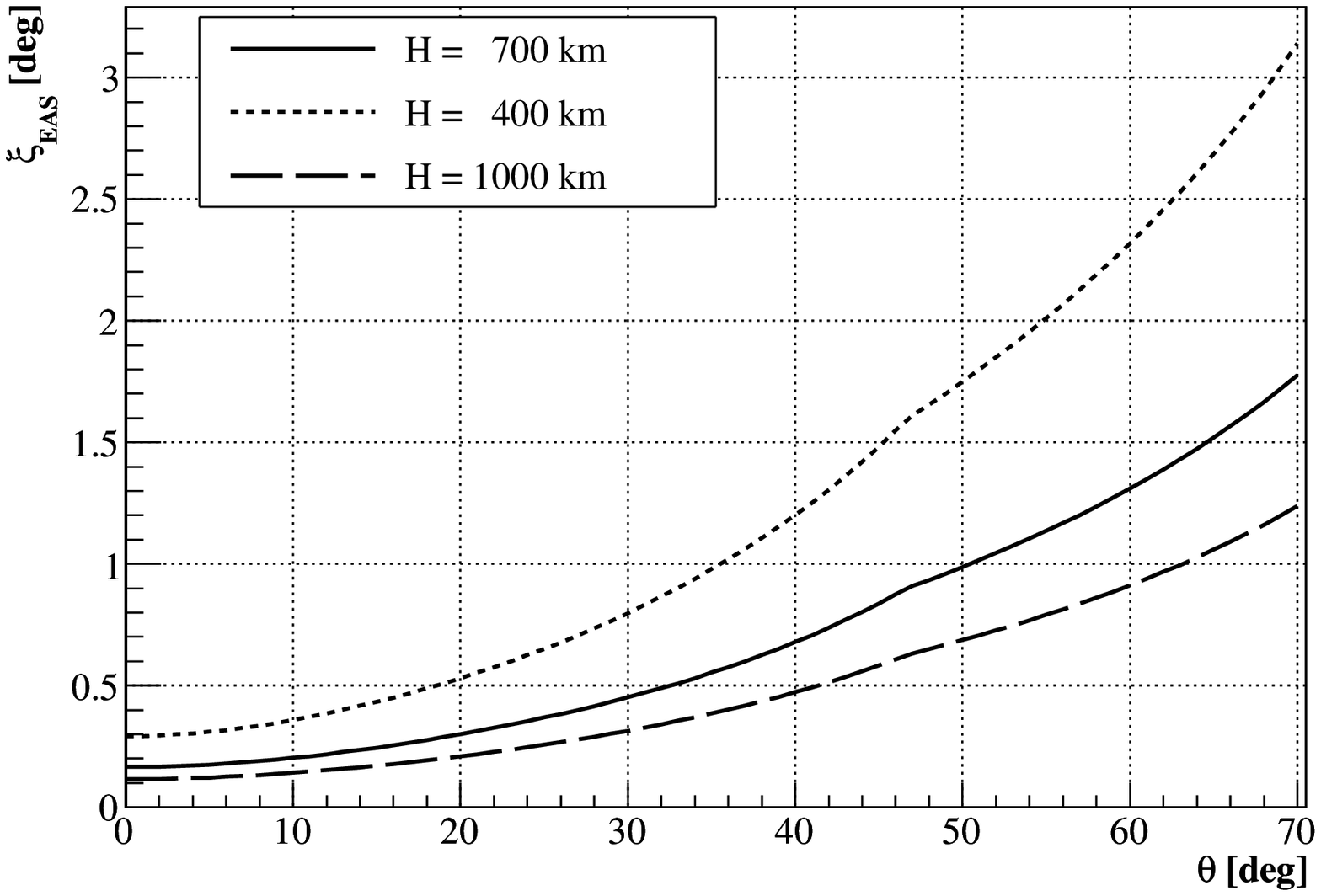} &    
\includegraphics[width=0.48\textwidth]{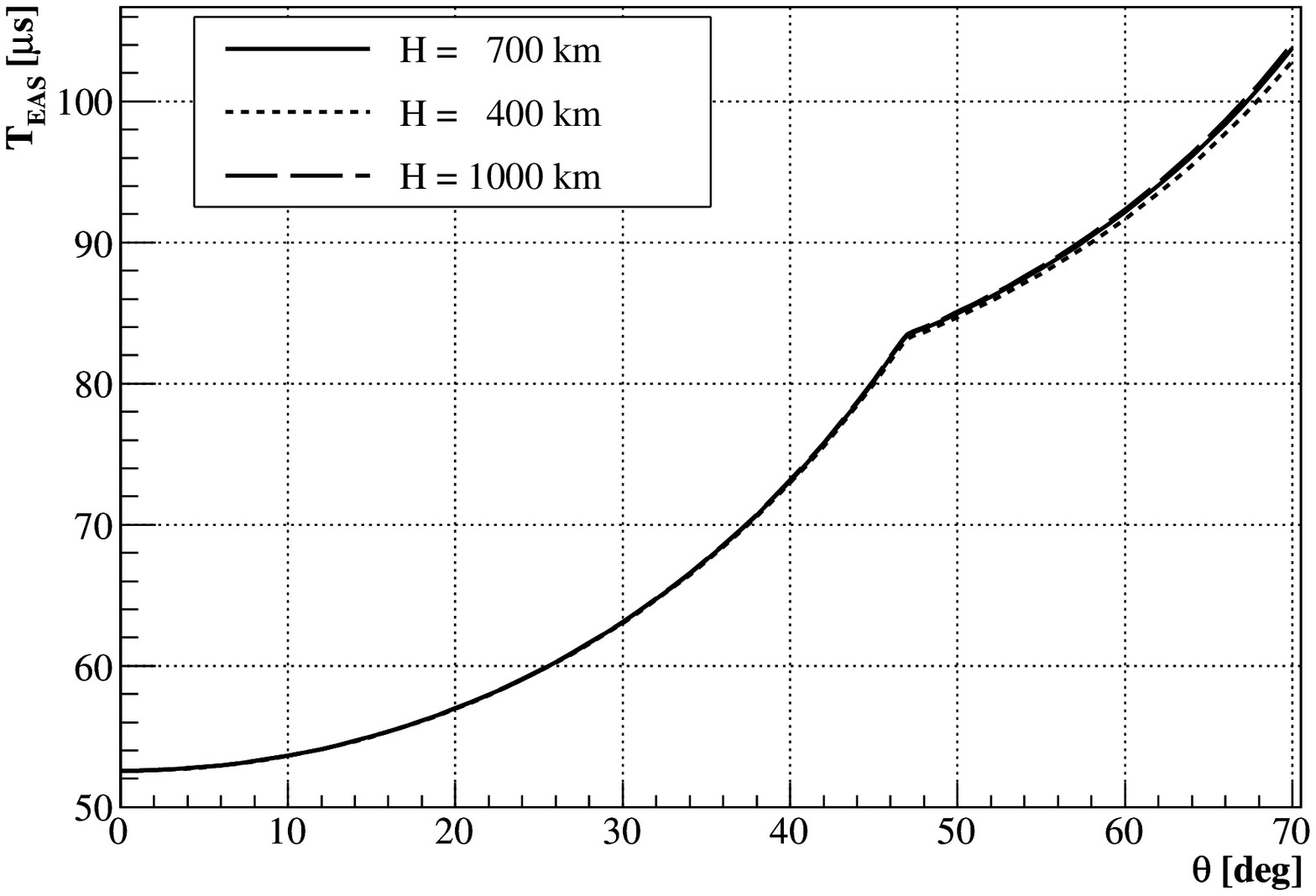}\\   
	(a)  &  (b) \\   
\end{tabular}   
		\caption{Angular extension (a) and time duration (b) of a $E=10^{19}$ eV proton \EAS image as a function of $\theta$ ($\psi_\mathrm{a}=90\degr$ and $\gamma=15\degr$). }   
		\label{fig:anglevstheta3h}   
\end{figure*}   
   
The Figures show that, typically, \mbox{$\EasAngle \lesssim 3\degr$} and $\EasTimeLength \lesssim 100\;\mu\text{s}$,   
justifying the assumptions of section~\ref{sec:GenAss}. Note also that,   
for $\theta\lesssim50\degr$, the \EAS are truncated because they  
hit the ground, as it is shown by the cusps in the   
figures. Nevertheless the truncation does not affect the reference \EAS (Table~\ref{tab:ReferenceConditions}).     
   
The time duration of the \EAS image is almost   
independent on both $H$ and $\gamma$ for a fixed \EAS geometry   
(see Figures~\ref{fig:anglevstheta3h}(b) and~\ref{fig:anglevstheta3gamma}(b)).
The angular length, for the same \EAS,   
will scale as \mbox{$\sim\cos\gamma/H$}. 

Some numerical values for the reference \EAS are given in section~\ref{sec:SignalEst}.
More plots can be found in~\cite{Pallavicini:2008wy}.

%--------------------------------------------------------------------------------   
\subsection{Aperture}   
\label{subsec:aperture}   
%--------------------------------------------------------------------------------   

%--------------------------------------------------------------------------------   
\subsubsection{Area observed at the Earth}   
%--------------------------------------------------------------------------------   
   
In the case of a nadir pointing telescope   
the geometrical area spanned by the \FoV at the surface of the Earth is    

\begin{equation}   
	A_0 = 2 \pi R_{\oplus}^2 \pton{ 1-\cos\beta_\mathrm{M} }     
\virgola  
\end{equation}   

in terms of $H$, the orbital height,   
$R_{\oplus}$, the Earth radius, $\Gmax$ the half-angle \FoV of the telescope,   
and $\beta_\mathrm{M}$, the angle at Earth center between nadir and the \FoV   
border

\begin{equation}
\beta_\mathrm{M} = \arcsin \left( \frac{R_{\oplus}+H}{R_{\oplus}}\sin\Gmax\right) - \Gmax
\end{equation}

The total atmosphere target mass, $M_\mathrm{a}$, can be estimated from $A_0 $ using the value   
of the vertical column density of air ($\sim 1033 \um{g/cm^2}$).   
Some values are given in Table~\ref{tab:Area} for different values of $H$ and
\Gmax.    

The expression for $A_0$ reduces, in the flat Earth   
approximation, to the well-known expression:    

\begin{equation}\label{eq:AreaFlatEarth}   
	A_0 = \pi H^2 \tan^2 \Gmax \punto   
\end{equation}

\begin{table}[htb]   
	\centering   
		\begin{tabular}{cccc} \hline   
			$H$         &  \Gmax      &  $A_0$  & $M_\mathrm{a}$  \\     \hline    
	                            &             & units of & units of                        \\
			$[\text{km}]$      &  degrees   & [\un[\sci{1}{5}]{km^2}] & [\un[\sci{1}{15}]{kg}] \\ \hline \hline   
			400       &  20  & 0.67  &   0.7 \\   
			700       &  20  & 2.07  &   2.1 \\   
			1000      &  20  & 4.25  &   4.4 \\   
			700       &  15  & 1.11  &   1.2 \\   
			700       &  25  & 3.43  &   3.5 \\ \hline   
		\end{tabular}   
	\caption{Area observed at the Earth and atmosphere mass target.}   
	\label{tab:Area}   
\end{table}

%--------------------------------------------------------------------------------   
\subsubsection{Instantaneous geometrical aperture}   
%--------------------------------------------------------------------------------   
   
The instantaneous geometrical aperture is defined as:   

\begin{equation}   
   \GeoAperture  \equiv \int_{A} \int_{\Omega} \hat{n}(\theta,\varphi)\cdot\hat{k}\diffl \Omega \diffl A   
    \virgola    
\end{equation}   

in terms of the normalized \EAS direction vector, $\hat{n}(\theta,\varphi)$, 
the normal unit vector to   
the Earth surface, $\hat{k}$, the target area, $A$, and the solid angle
$\Omega$.   
For \EAS with $0\le \varphi \le 2\pi$ and $0\le\theta\le\frac{\pi}{2}$,
the instantaneous geometrical aperture reduces to

\begin{equation}   \label{eq:GeoApertureSimple}
    \GeoAperture = \pi A_0   
    \punto    
\end{equation}   
   
The instantaneous geometrical aperture is related to the rate of observed \EAS
by the relation   

\beq   
        \deriv{N_\mathrm{events}}{t} = \GeoAperture J(E\geq E_\mathrm{TH}) \virgola   
\eeq   
where $J$ is the flux of primary UHECP.
   
Equation~\ref{eq:GeoApertureSimple} implies that
in order to reach $\GeoAperture \gtrsim \un[10^6]{km^2}\un{sr}$ it has to be
$\beta_\mathrm{M} \gtrsim 2.86\degr$.
 The choice $H=\un[700]{km}$ and $\Gmax=25\degr$   
satisfies this requirement.

%--------------------------------------------------------------------------------   
\subsubsection{Effective aperture}   
%--------------------------------------------------------------------------------   

The effective aperture of a nadir pointing telescope,   
when considering all the \EAS which reach the ground inside the \FoV at any
zenith angle, 
is approximately given by the relation:   
\begin{equation}   
\begin{split}   
	\EffAperture &=   
	\eta_\mathrm{o} \eta_\mathrm{c} \pton{ 1-\tau_{\mathrm{dead}}} \GeoAperture(\Gmax, H) \\   
   &\approx	2 \eta_\mathrm{o} \eta_\mathrm{c} \pton{ 1-\tau_{\mathrm{dead}} } \pi^2 R_{\oplus}^2 \pton{ 1-\cos\beta_\mathrm{M} }   
   \virgola   
\end{split}   
\end{equation}   
in terms of the orbital height, $H$, the half-angle \FoV $\Gmax$,    
the observational duty cycle $\eta_\mathrm{o}$, the dead time $\tau_\mathrm{dead}$    
and the factor $\eta_\mathrm{c}\sim 0.5$,    
quantifying the effect of real cloud coverage on the \EAS detection efficiency.   
This expression gives the asymptotic aperture,   
when the total detection efficiency of the \EAS is equal to one.   
   
%--------------------------------------------------------------------------------   
\subsubsection{Tilting of the telescope}\label{sec:tilting}   
%--------------------------------------------------------------------------------   
   
The instantaneous geometrical aperture can increase if the telescope is   
tilted with respect to the local nadir by some angle $\TiltAngle$.    
   
In this case, the geometrical solution to the problem of   
finding the intersection area of the circular \FoV cone with the spherical   
Earth surface is not trivial at all. The easiest way to estimate the aperture is therefore to use a simple   
Monte-Carlo integration.   
   
\begin{figure*}[htb]   
	\centering   
\begin{tabular}{cc}   
		\includegraphics[width=0.48\textwidth]{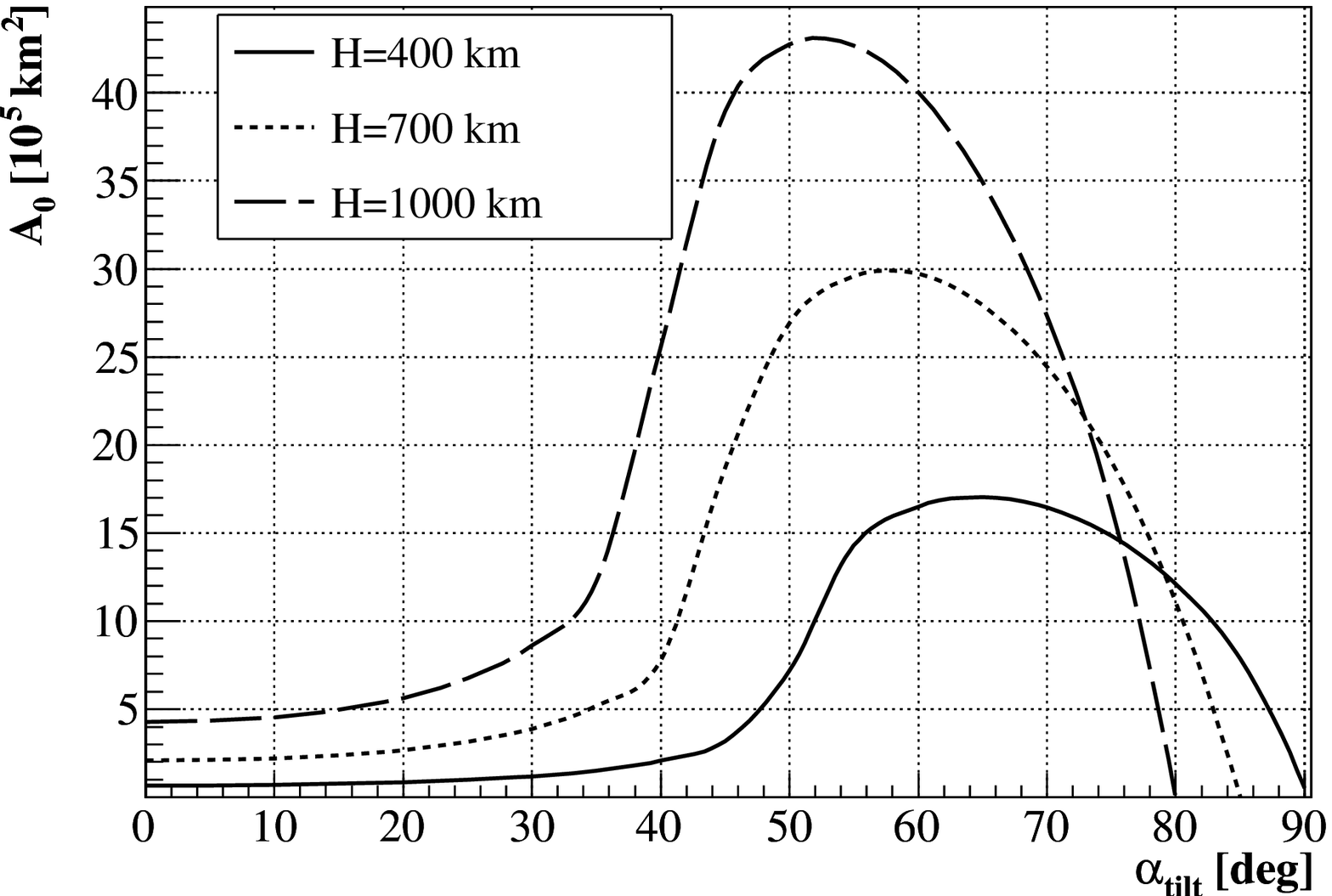} &   
\includegraphics[width=0.48\textwidth]{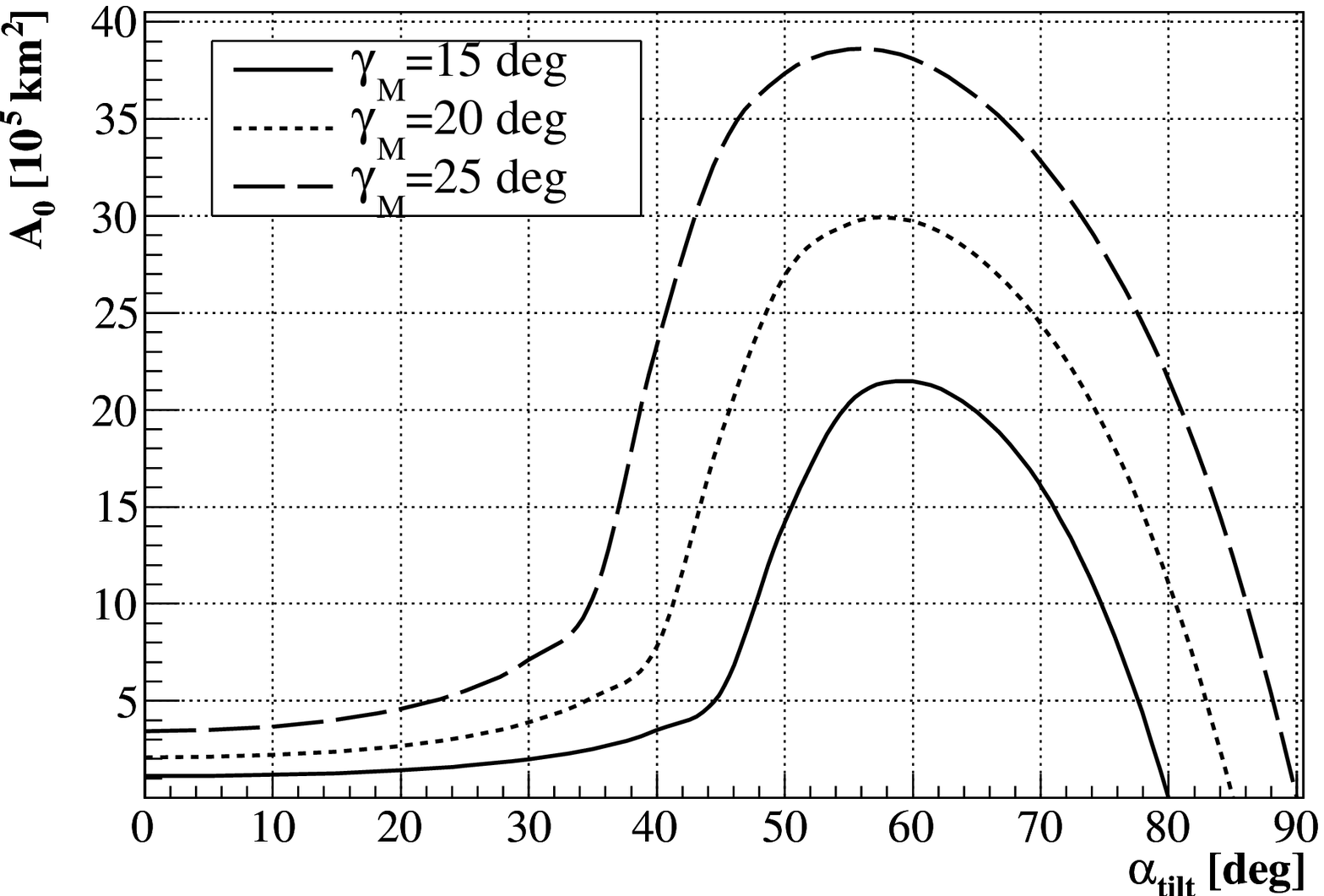} \\   
(a) & (b) \\   
\end{tabular}   
	\caption{Observed $A_0$ versus   
		$\TiltAngle$. (a): for $\Gmax = 20\degr$;  (b): for $H = \un[700]{km}$.}   
	\label{fig:tiltmulti}   
\end{figure*}

The Monte-Carlo results for the intersection area, as a function of $\TiltAngle$, are shown   
in Figure~\ref{fig:tiltmulti} for different values of $H$ and \Gmax.    
   
It should be   
noted that the corresponding horizon angle is    
\begin{gather}
\beta_\mathrm{hor} = \arcsin\left(\frac{R_{\oplus}}{R_{\oplus}+H}\right)
\\
\approx    
\begin{cases}   
	 \hfill 70\degr & \text{at}\: \hfill H=400\um{km} \\   
	 \hfill 64\degr & \text{at}\: \hfill H=700\um{km} \\   
	 \hfill 60\degr & \text{at}\: \hfill H=1000\um{km}    
\end{cases}   
\virgola   
\end{gather}   
which has to be taken into account when computing the maximum possible tilting   
angle.   
   
The instantaneous geometrical aperture, in tilted mode,
can increase up to   
a factor $(3 \div 5)$, with respect to the case of no tilt.   
   
The main drawback of tilting is that the atmospheric absorption
increases at the far extreme of the \FoV (see Table~\ref{ta:AtmoTransm}).   
This means that the effective energy threshold in the far part of the   
\FoV increases.   
Moreover the \FoV is non-uniform and the angular resolution at the far   
extreme becomes worse than at nadir.
The large \FoV of the optics, together with the large atmospheric   
target observed, would require an excellent stray-light control for a   
tilted telescope due to   
the large amount of stray light entering the \FoV.   
Tilting, together with the large \FoV, would also strongly affect the   
duty cycle, by a factor which cannot be easily estimated semi-analytically,    
as the large area observed at the Earth would more often include   
day-time areas and sources of stray light.   
   
On the other hand increasing the orbit height seems to be a more effective method   
to increase the instantaneous geometrical aperture while minimizing the   
drawbacks mentioned above.   
In fact one would have no significant losses from the increased atmospheric transmission   
and a more uniform \FoV.   
A highly elliptical orbit is also an option.

%--------------------------------------------------------------------------------   
\subsubsection{Duty cycle}   
%--------------------------------------------------------------------------------   
   
It is not easy to estimate the duty-cycle of a space-based telescope   
without a dedicated preliminary measurement of the real background seen from the
real satellite orbit.

The observational duty cycle is driven by the acceptable background level   
and it is therefore dependent on the range of energies of the \EAS under study.   
However to allow for background dependent observations implies a very precise knowledge    
of the telescope sensitivity as a function of the background level.

The major limitation to the telescope observational duty cycle comes from Sun light and Moon light.    
The fraction of time  the telescope is unable to operate depends on the orbit and, during moon-time, on the   
maximum background rate such that data taking, that is online background
subtraction, is possible.   

For an orbital height $H\sim \un[400]{km}$, an orbit inclination of $\sim   
50\degr$ and requiring both the Sun and the Moon to be safely below the   
horizon, the average estimated duty cycle is $\eta_\mathrm{o}\sim 13\%$. 
If one accepts an additional background flux of   
\mbox{$B_\mathrm{ML} \approx \un[100]{ph}\um{m^{-2}}\um{ns^{-1}}\um{sr^{-1}}$}, due to   
Moon light, the estimated duty cycle is \mbox{$\eta _\mathrm{o}\sim 19\%$}~\cite{Berat}.   
   
The duty cycle may be also influenced by man-made or other natural sources (see section~\ref{subsec:NoiseBackground}).

%--------------------------------------------------------------------------------   
\subsection{Pixel size}\label{sec:pixelsize}   
%--------------------------------------------------------------------------------   
   
The pixel size, $\delta$, is driven by the scientific requirements and
constrained by the available technology and resources.    

The number of signal   
photons divided by the number of random background photons on the   
pixel roughly scales as $S/B \sim 1/\Delta\alpha $    
(where $\Delta\alpha$ is the angular pixel size corresponding to $d$, see equation~\eqref{eq:angulargran})   
for a pixel much larger than the \EAS   
track width on the focal surface (which basically depends on the optics PSF)    
while it saturates to a constant for a small enough pixel size.

The pixel size also affects
the angular resolution and    
the \Xmax resolution.   
   
Due to the relatively small \EAS   
transverse dimensions, the width of the \EAS image on the focal surface will   
be determined by the PSF only. 

A pixel size roughly the same size of the optics PSF turns out to be,   
usually, a good compromise as it allows to exploit the optics performance.
On the contrary, a pixel size much smaller than the optics PSF requires a much
larger number of detector channels and allows to obtain a sub-pixel resolution,
by fitting the known incident photon distribution, 
only in case the number of incident photons is large.

An approximate and simplified analysis, leading to determine the required pixel   
size starting from the angular resolution, is presented in Section~\ref{sec:angularres}.    
   
%-------------------------------------------------------------------------------   
\subsection{Angular resolution}\label{sec:angularres}   
%------------------------------------------------------------------------------- 

Any \EAS will be seen as a point moving inside the \FoV with a kinematics   
(direction and angular velocity) determined by the \EAS direction relative to   
the line-of-sight from the telescope to the   
\EAS instantaneous position.   
The direction of the \EAS velocity vector, as seen by the telescope,   
can be decomposed into two perpendicular components: the one parallel to the line-of-sight    
and the other one lying in the plane perpendicular to it.     
The \EAS develops approximately at the speed of light and its distance can be   
considered as a known value, in the case of a space-based experiment.   
Both components can be reconstructed from    
the two-dimensional image on the focal surface plus the timing information.   
   
Using the reference conditions of Table~\ref{tab:ReferenceConditions} and exploiting the   
Gamma-like shape of the \EAS longitudinal profile,   
the relation between the observed \EAS angular length,   
$\EasAngle$, and the standard   
deviation of the observed longitudinal photon distribution along the \EAS image,   
$\sigma_{\xi}(\mathcal{N})$, turns out to be $\EasAngle = 5\sigma_{\xi}$    
(for $\mathcal{N}=100$), as discussed in section~\ref{sec:vislength}.   
   
The determination of angular resolution of the two components, perpendicular and parallel to the line-of-sight,   
are presented respectively in Sections~\ref{sec:angrperp} and~\ref{sec:angrpar}.   

Note that the following elementary analysis ignores   
the effect of the background, which will make the angular resolution worse.
On the other hand, the use of the diffusely reflected \Cherenkov flash, if implemented, might improve the   
angular resolution.   
   
%-------------------------------------------------------------------------------   
\subsubsection{Angular resolution perpendicular to the line-of-sight}\label{sec:angrperp}   
%-------------------------------------------------------------------------------   
   
The expected angular resolution $\Delta \beta_{\perp}$ on the \EAS direction   
perpendicular to the line-of-sight is readily estimated by    
a linear fit. The error on the angle can be calculated from the 
standard errors relations for the least squares fit to a straight line with equal errors in both
variables as~\cite{bib:petroLSF}:   

\begin{equation}   \label{eq:AngResPerp}
	\Delta \beta_{\perp}(\mathcal{N}) = \frac{\Delta\alpha }{\sqrt {12}}\frac{1}{\sigma_\xi (\mathcal{N})}\frac{1}{\sqrt{\mathcal{N}}}  
	\virgola   
\end{equation}

where $\sigma_\xi (\mathcal{N}=100)\approx \EasAngle(\mathcal{N}=100)/5$ (see
Section~\ref{sec:vislength})
and $\Delta\alpha$ is the angular pixel size.

Note that the above result is consistent with the naive evaluation:   

\begin{equation}   
	\Delta \beta_{\perp} \approx \frac{\Delta\alpha}{\EasAngle}\frac{1}{\sqrt {\mathcal{N}}}   
	\punto   
\end{equation}

%--------------------------------------------------------------------------------   
\subsubsection{Angular resolution parallel to the line-of-sight}\label{sec:angrpar}   
%--------------------------------------------------------------------------------   

The relation between the observed angular velocity \EasAngularVelocity and the   
angle $\beta_{\parallel}$ between the \EAS velocity vector and the line-of-sight   
is the well known relation~\cite{Sommers:1995dm,Baltrusaitis:1985mx}:   
   
\begin{equation}   
    \EasAngularVelocity =   
    \frac{c}{D} \pton{ {\frac{1-\cos \beta_{\parallel} }{\sin \beta_{\parallel} }} } =    
    \frac{c}{D}\tan \pton{ {\frac{\beta_{\parallel}}{2}} }   \virgola
\end{equation}   
where $c$ is the speed of light and $D$ is the distance of the \EAS.

In the   
present case, one assumes that the \EAS develops in the lower layers of the   
atmosphere, within $\sim\un[15]{km}$ from the ground, so that $D$ is   
approximately known (the relative error is $\Delta D / D \lesssim 0.05 $).    
   
The error on the angle $\beta_{\parallel}$ can be estimated from the equation
$\alpha(t)=\EasAngularVelocity t\:$ using the standard errors relations for the least squares fit to a straight line with equal errors in both
variables as~\cite{bib:petroLSF}:   

\beq   
\begin{split}   \label{eq:AngResPara}
\Delta \beta_{\parallel}(\mathcal{N}) 
& = 
 \frac{K}{\sqrt{\mathcal{N}}} \frac{1}{\sqrt{\left(\frac{\sigma_t(\mathcal{N})}{\Delta\tau}\right)^2
    +\left(\frac{\sigma_\xi(\mathcal{N})}{\Delta\alpha}\right)^2} }
\\   
& \approx \frac{1}{\sqrt{\mathcal{N}}}    
  \frac{1}{\sqrt{\left(\frac{\sigma_t(\mathcal{N})}{\Delta\tau}\right)^2 +\left(\frac{\sigma_\xi(\mathcal{N})}{\Delta\alpha}\right)^2} }
\virgola
\\
K &\equiv  \frac{2}{\sqrt{12}} \frac{D \Delta\alpha}{c\Delta\tau}    
 \frac{\left[ 1 + \left( \frac{c \Delta\tau}{D \Delta\alpha}\right)^2 \tan^2
     \left(\frac{\beta_{\parallel}}{2} \right) \right]}{\left[ 1+ \tan^2
     \left(\frac{\beta_{\parallel}}{2} \right)\right]} 
\virgola   
\end{split}   
\eeq   

where $\Delta\tau$ is the time resolution of the telescope, while 
\mbox{$\sigma_\xi (\mathcal{N}=100)\approx \EasAngle(\mathcal{N}=100)/5$} and
\mbox{$\sigma_t (\mathcal{N}=100)\approx \EasTimeLength(\mathcal{N}=100)/5$} (see
Section~\ref{sec:vislength})
and $\Delta\alpha$ is the angular pixel size.
  
%--------------------------------------------------------------------------------   
\subsubsection{Requirements deriving from the angular resolution}   
%-------------------------------------------------------------------------------- 

In the reference conditions of Table~\ref{tab:ReferenceConditions}, a \FoV
granularity of $\Delta\alpha \sim 0.1\degr$ is required, from
equation~\ref{eq:AngResPerp}, in order to reach
an angular resolution of the order of $\Delta \beta_{\perp}\sim 1\degr$.
Moreover, in order to reach 
an angular resolution of the order of $\Delta \beta_{\parallel}\sim1\degr$, 
a time resolution $\Delta\tau = \un[2.5]{\mu s}$ is required, from
equation~\ref{eq:AngResPara}.

The above results lead to two similar values of the angular resolutions
$\Delta \beta_{\perp}$ and  $\Delta \beta_{\parallel}$, which can be written,
neglecting the soft (roughly logarithmic) dependence on $\mathcal{N}$ of 
$\sigma_\xi$ and $\sigma_t$, as:

\begin{equation}
        \Delta \beta_{\perp}(\mathcal{N}) \approx \frac{0.14}{\sqrt{\mathcal{N}}}  
        \spazio
        \Delta \beta_{\parallel}(\mathcal{N}) \approx \frac{0.12}{\sqrt{\mathcal{N}}}   
\punto
\end{equation}

\begin{figure*}[htb]   
	\centering   
\begin{tabular}{cc}   
		\includegraphics[width=0.48\textwidth]{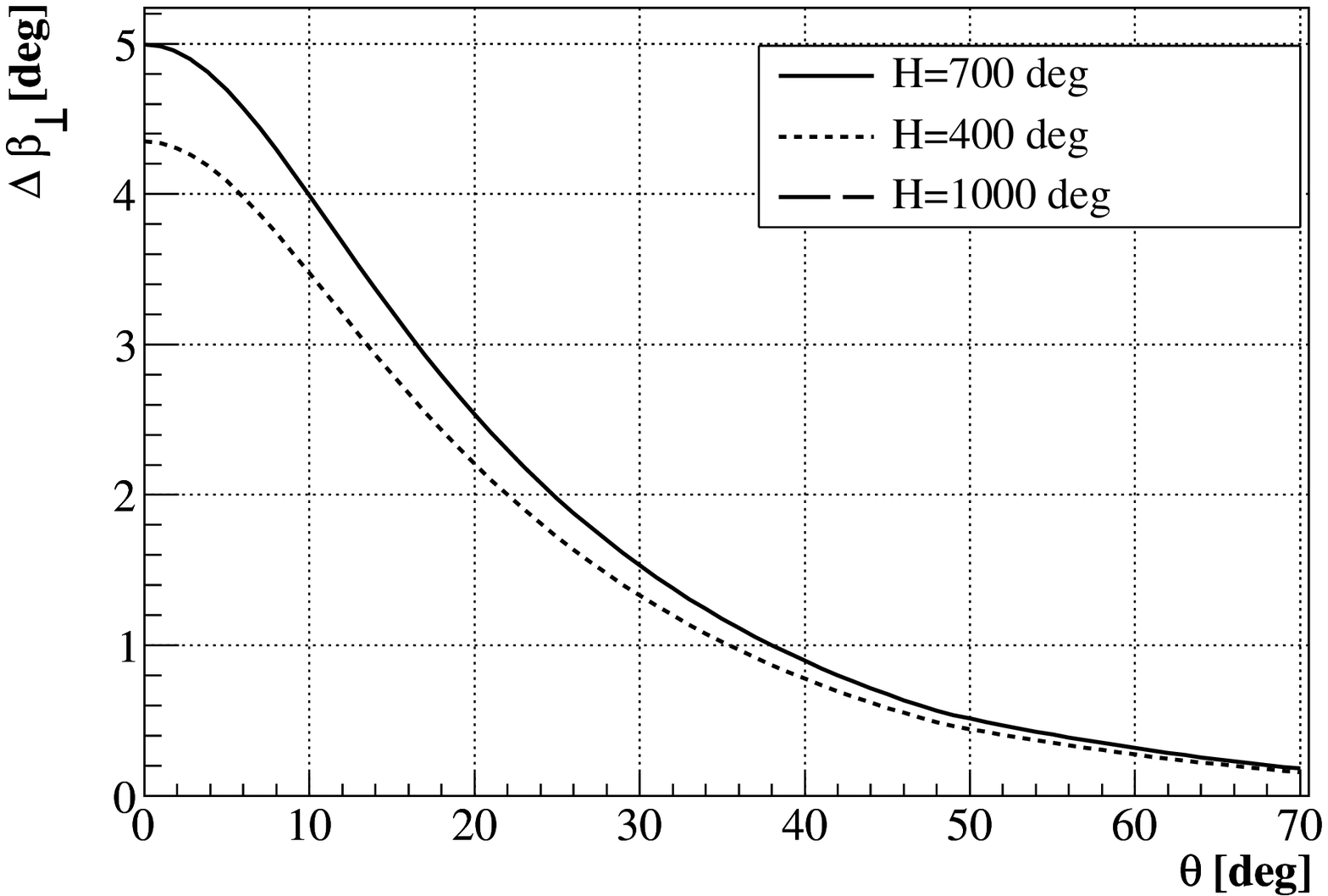} &   
\includegraphics[width=0.48\textwidth]{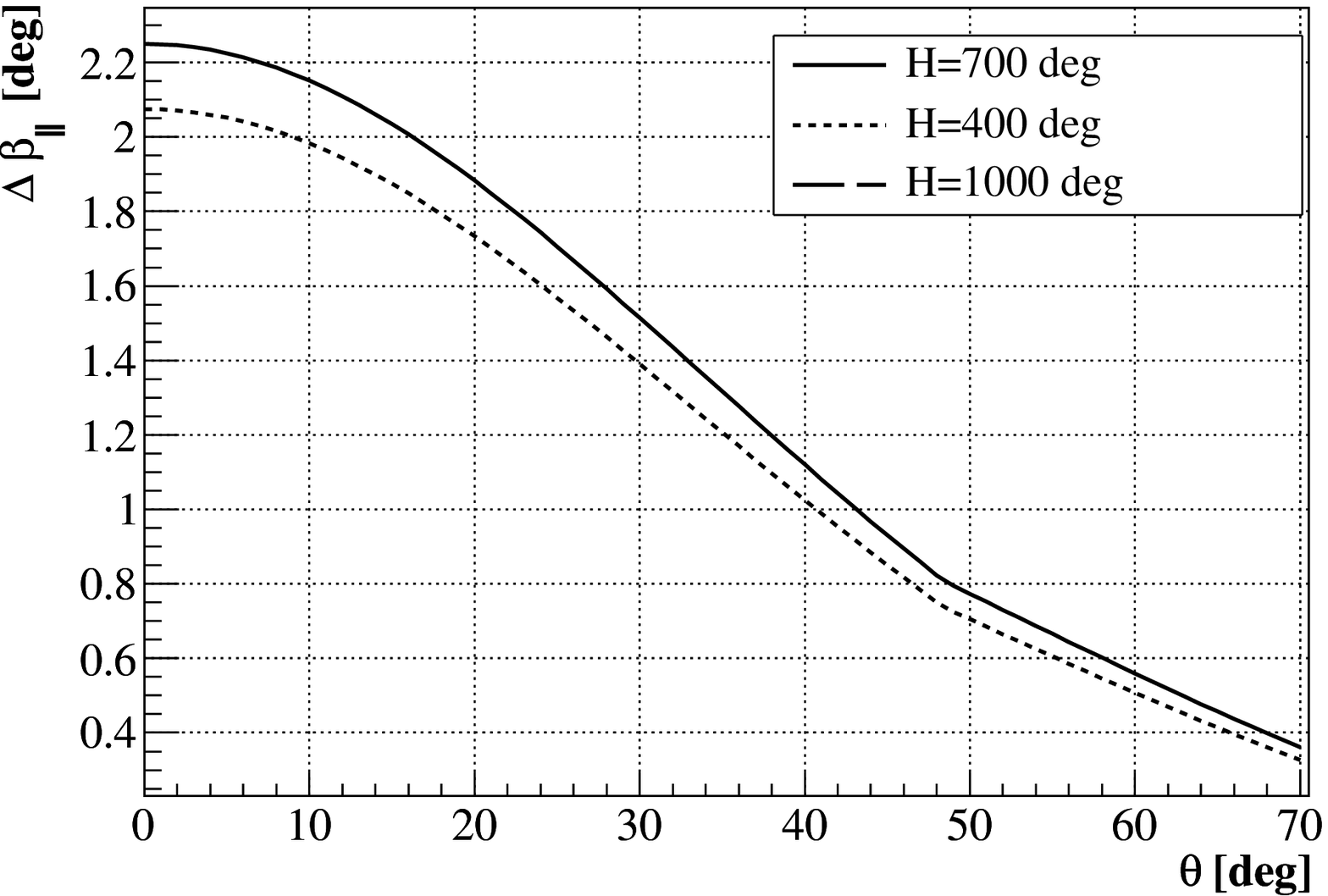} \\   
(a) & (b) \\   
\end{tabular}   
		\caption{Angular resolution $\Delta \beta_{\perp}$ (a) and $\Delta \beta_{\parallel}$ (b) as a   
		function of $\theta$ in the reference conditions of Table~\ref{tab:ReferenceConditions}. Solid line:   
		$H=700\;\text{km}$; dotted line: $H=400\;\text{km}$ (the   
		line for $H=1000\;\text{km}$ is very close to the first   
		one). $\Delta\tau=\un[2.5]{\mu s}$, $\Delta \alpha=0.1\degr$.  }
	\label{fig:AngRes}   
\end{figure*}

Other numerical estimates are given in section~\ref{sec:AngResEst}.    
The angular resolution $\Delta \beta_{\perp}$ as a function of $\theta$,   
is shown in Figure~\ref{fig:AngRes}(a).    
The angular resolution $\Delta \beta_{\parallel}$ as a function of   
$\theta$, is shown in Figure~\ref{fig:AngRes}(b).

%--------------------------------------------------------------------------------   
\subsection{\Xmax resolution}   
%--------------------------------------------------------------------------------    

The equation~\ref{eq:slantdepth},
defines the slant depth $X(\ell)$ as a function of the   
coordinate along the \EAS, $\ell$, and the atmospheric density, $\rho(h(\ell))$.
Therefore one can derive an estimate of the angular resolution required to
obtain any given \Xmax resolution by applying the equation at an altitude 
$h \approx 6 \um{km}$, where the \Xmax of the reference \EAS is located (see
figure~\ref{fig:hmax}):

\begin{equation}   
    \DD{X} \simeq   \rho(h(\ell))  \DD{\ell} 
\end{equation}   

In the reference conditions of Table~\ref{tab:ReferenceConditions}, it is
\mbox{$\Delta\ell\approx\un[1.2]{km}$},
for $\Delta\alpha \sim 0.1\degr$ (as determined in the previous section), so 
that: $ \DD{\Xmax} \simeq 90 \um{g/cm^2} $.
This shows that, in the reference conditions and neglecting the geometrical and kinematical details of the   
\EAS development, one can estimate that in order to reach the required
\Xmax resolution of $ \DD{\Xmax} \simeq 50 \um{g/cm^2} $
an angular granularity of the order of $\Delta\alpha \sim 0.05\degr$ is required.

%--------------------------------------------------------------------------------   
\subsection{Background and noise}   
\label{subsec:NoiseBackground}   
%--------------------------------------------------------------------------------   

A wide variety of background sources can affect
the detection of \UHECP from space. For every source
the luminosity and variability in space and time
must be studied and the data must be related to
the local weather conditions.

Actually the sources of background to the \EAS
observation can be roughly divided into
four main categories.

There are almost-continuous natural night sky diffuse
and slowly varying sources, uniformly distributed in the \FoV. The main
component is the atmospheric night-glow~\cite{bib:UVng}. Also the moon-reflected
sunlight contributes significantly during the
full moon time.

There are ground and man-made sources, localized
in the \FoV, such as city lights or other luminous
sources slowly moving with respect to the
satellite speed projected on ground (airplanes, cars, ...).
The sources of this class could blind the detector in the corresponding region of
the \FoV. The knowledge of their position
in the \FoV and their emission
properties in the observational wavelength
range will be helpful to limit their effect on
the instrument duty cycle.

Transient Luminous Events (TLE)~\cite{bib:tle} in the lower atmosphere
(lightnings) and in the upper atmosphere (red sprites, elves
and blue jets) were discovered
in the late eighties and are still poorly known. Recent
measurements~\cite{bib:tatianaflashes} have shown the existence
of extremely powerful flashes in the near
UV which may be related with the physics of the upper
atmosphere. These luminous flashes might threaten
the integrity of the photo-detector, if not properly
taken into account. Therefore, the design of the
orbit must optimize the duty cycle and
minimize the aging effects,
by avoiding the regions where the TLE are
more frequent.

Finally there are other luminous
events: satellites and debris in the sun light,
meteorites, low energy cosmic radiation, satellite
and satellite glow. 

In this paper, looking for the minimal necessary requirements
for the observation, 
only the continuous random background is taken into account, since
it can be directly measured and subtracted. 
All the other background sources require a different handling and their
space-time structure, distinct from the one of the \EAS, must be exploited to
reject them.   
   
The total random background rate intercepted by the telescope    
(on the whole entrance pupil and full \FoV) is:   

\beq   
	\nu_\mathrm{tot}^\mathrm{b} = B A_\mathrm{EP} \pi \sin^2 \Gmax    
\punto   
\eeq   
   
A typical value for the    
total rate of background hits detected is of the order of $\un[10]{THz}$.
See Section~\ref{sec:NightGlowEst} for some numerical estimates.  

The intrinsic noise generated by all the parts of the experimental telescope    
has to be kept well below the physical background level, due to the faintness of the air scintillation signal. 
It is therefore required to be less   
than a few GHz (over the whole photo-detector) in order to be negligible with   
respect to the background.    
Note that this requirement should also include the   
stray-light coming from lack of light-tightness of the telescope.   
   
The online subtraction of the random background is mandatory as its rate is not negligible   
with respect to the rate of the air scintillation signal, especially for \EAS near the energy threshold.   
A continuous monitoring of the average background on a pixel-by-pixel basis,
is therefore unavoidable to subtract it and to lower the energy threshold.
   
Many experiments have measured the intensity of the random background from space:   
BABY~\cite{baby}, NIGHTGLOW~\cite{nightglow},    
Arizona-Airglow~\cite{arizona} and the Universitetsky-Tatiana microsatellite~\cite{tatiana}.    
However, in order to devise a   
method for online background subtraction, it is necessary to have a finer   
characterization of the space-time behavior of the background   
on the space-time scale of the \EAS development: $ \DD{\alpha} \simeq 0.1\degr$   
and $ \DD{t} \simeq \um{\mu s}$.   
This characterization should include measurements   
along different directions from nadir to cope with off-field background~\cite{spacepart06}.   
These measurements are relevant due to the large \FoV, especially with a tilted telescope.   
   
%--------------------------------------------------------------------------------   
\section{Conclusions}   
\label{sec:Conclusions}   
%--------------------------------------------------------------------------------   
   
The design of a space-based telescope aimed at the observation of the \EAS
produced by \UHECP is a very challenging task, particularly because of the small
photon signal, the large photon background, the harsh space environment and the
limited resources of a space mission.  
It is therefore very important to
optimize the performance of the telescope, after defining the scientific
requirements, and set safe design margins at the beginning of the concept study
itself.

In this paper we have presented a set of 
analytical formulas and semi-analytical results
that might be useful for the design of a future space-based experiment. 
These formulas and results allows one to set the approximate values of the
design parameters, starting from given scientific requirements.
These design parameters can be refined, afterwards, with detailed, as well as
heavy and time consuming, full Monte-Carlo simulations.
  
The formulas and results were used to estimate the expected performance of
realistic satellite configuration, the \SEUSO mission proposed to ESA in the
framework of the Cosmic Vision program 2015-2025~\cite{cosmicvision}.

We believe that before starting such a large and challenging free-flyer mission,
a number of intermediate steps are needed, including preliminary measurements
of the background and some sort of technology demonstrator.  
The envisaged JAXA
JEM-EUSO~\cite{jemeuso} mission might well play the role of a path-finder.
   
%--------------------------------------------------------------------------------   
\section*{Acknowledgments}   
\addcontentsline{toc}{section}{Acknowledgments}   
%--------------------------------------------------------------------------------   
   
The authors wish to thank the members of the \EUSO Collaboration and in particular   
Osvaldo Catalano (INAF/IASF-Palermo),   
Didier Lebrun (LPSC, Grenoble),   
Piero Mazzinghi (INOA, Firenze),     
Sergio Bottai (INFN, Firenze)   
Andrea Santangelo (IAAAA, T\"ubingen),   
Dmitry Naumov (JINR, Dubna),   
Lloyd Hillman\symbolfootnote[2]{Deceased.} (University of Alabama, Huntsville),   
Yoshi Takahashi$^\dagger$ and   
John Linsley$^\dagger$   
for many useful discussions and suggestions.   
   
The pioneering work~\cite{arisaka} of Katsushi Arisaka (UCLA, Los Angeles) is acknowledged.   
   
All this would have not been possible without the boost of Livio Scarsi$^\dagger$.   

The many constructive comments/criticisms of two anonymous referees are acknowledged.
   
\appendix 
  
%-------------------------------------------------------------------------------   
\section{Some order of magnitude estimates}\label{sec:Estimates}   
%-------------------------------------------------------------------------------   
   
%-------------------------------------------------------------------------------   
\subsection{Air scintillation signal}\label{sec:SignalEst}   
%-------------------------------------------------------------------------------   

Let us consider the reference conditions of Table~\ref{tab:ReferenceConditions}
and an ideal telescope optics 
($ \mathcal{E}^\prime_\mathrm{O} (\gamma) = A_\mathrm{EP}\cos\gamma $).   
   
The signal time-integrated irradiance reaching the telescope is given in
Table~\ref{tab:irradiance}, for different values of $\theta$ and $H$, the values
of the other parameters being as in Table~\ref{tab:ReferenceConditions}.    
Note that the irradiance scales as \mbox{$\sim\left(\cos\gamma/H\right)^2$}.   
   
\begin{table}[htb]   
\begin{center}   
\begin{tabular}{c|c|c|c} \hline   
$\diffl{\mathcal{N}}/\diffl{A}\:[\un{ph}\un{m^{-2}}]$ & $\theta=30\degr$ & $\theta=50\degr$ & $\theta=70\degr$ \\ \hline \hline   
$H=\un[400]{km}$  & 40 & 70 & 180 \\   
$H=\un[700]{km}$  & 15 & 25 & 60 \\   
$H=\un[1000]{km}$  & 5 & 10 & 30 \\ \hline   
\end{tabular}   
\caption{Signal time-integrated irradiance, $\diffl{ \mathcal{N} }/\diffl{A}$,
  in the reference conditions of Table~\ref{tab:ReferenceConditions}.}\label{tab:irradiance}   
\end{center}   
\end{table}

The detection of such an \EAS obviously requires a large entrance pupil of many
square meters, in order to get a number of detected photons of the order of a
few hundreds as the total efficiency (optics, filters and photo-detector) is
$\varepsilon_\mathrm{tot} \approx 0.1 $.

The typical angular length and the time duration of the \EAS image are given
respectively in Tables~\ref{tab:anglength} and~\ref{tab:duration}.   

\begin{table}[htb]   
\begin{center}   
\begin{tabular}{c|c|c|c} \hline   
$\EasAngle\:[\text{deg}]$ & $\theta=30\degr$ & $\theta=50\degr$ & $\theta=70\degr$ \\ \hline \hline   
$H=\un[400]{km}$  & 0.8 & 1.7 & 3.1 \\   
$H=\un[700]{km}$  & 0.4 & 1.0 & 1.8 \\   
$H=\un[1000]{km}$  & 0.3 & 0.7 & 1.2 \\ \hline   
\end{tabular}   
\caption{Angular length \EasAngle in the reference conditions of Table~\ref{tab:ReferenceConditions}.}\label{tab:anglength}   
\end{center}   
\end{table}   
   
\begin{table}[htb]   
\begin{center}   
\begin{tabular}{c|c|c|c} \hline   
$\EasTimeLength\:[\un{\mu s}]$ & $\theta=30\degr$ & $\theta=50\degr$ & $\theta=70\degr$ \\ \hline \hline   
$H=\un[400]{km}$  & 60 & 84 & 102 \\   
$H=\un[700]{km}$  & 60 & 84 & 103 \\   
$H=\un[1000]{km}$  & 60 & 84 & 104 \\ \hline   
\end{tabular}   
\caption{Time duration \EasTimeLength in the reference conditions of Table~\ref{tab:ReferenceConditions}.}\label{tab:duration}   
\end{center}   
\end{table}

%-------------------------------------------------------------------------------   
\subsection{Requirements on the optical triggering efficacy}\label{sec:TheoOptRequ}   
%-------------------------------------------------------------------------------   
   
Following the discussion in Section~\ref{sec:NumPh}, the    
required overall photo-detection efficacy in the reference conditions of Table~\ref{tab:ReferenceConditions}   
can be calculated   
\[   
 \PDEfficacy(\gamma=15\degr) \approx \mathcal{N}\left(\deriv{ \mathcal{N} }{A}\right)^{-1}  \virgola   
 \]    
where $\mathcal{N}\simeq 100$ at the energy threshold. Using the values of the time-integrated irradiance given in the   
previous section it is $\PDEfficacy(\gamma=15\degr)\approx \un[4]{m^2}$   
when observing at $H=\un[700]{km}$, $\PDEfficacy(\gamma=15\degr)\approx \un[1.5]{m^2}$   
at $H=\un[400]{km}$ and $\PDEfficacy(\gamma=15\degr)\approx \un[10]{m^2}$    
at \mbox{$H=\un[1000]{km}$}.   
   
Assuming conservatively $\varepsilon_\mathrm{f}\PDE \sim 0.1$, 
the optical triggering efficacy is obtained from the equation~\eqref{eq:telefficacy},   

\begin{equation}\label{eq:OptEfficacy:Req}
\begin{split}   
    \mathcal{E}_\mathrm{O}'(\gamma=15\degr) &\approx \frac{\PDEfficacy}{\varepsilon_\mathrm{f}\PDE} \approx \\
&\approx   
    \begin{cases}   
     \hfill \un[15]{m^2}  &\text{at}\: \hfill H=\un[400]{km} \\   
     \hfill \un[40]{m^2}  &\text{at}\: \hfill H=\un[700]{km} \\   
     \hfill \un[100]{m^2} &\text{at}\: \hfill H=\un[1000]{km} \\   
    \end{cases}      
\punto
\end{split}   
\end{equation}

The optical triggering efficacy at $\gamma=0\degr$, that is the entrance pupil area,   
is not only the effective photon collection area but it is also an estimate of the   
physical area of the optics, actually an optimistic one. Therefore it translates
into a requirement on the minimum size of the telescope.     
   
For an ideal optics, such that the optical triggering efficacy is   
\mbox{$\mathcal{E}_\mathrm{O}'(\gamma)=A_\mathrm{EP}\cos\gamma\,$}, it is   

\begin{equation}\label{eq:ADest}   
    \begin{split}   
        A_\mathrm{min}&\approx\frac{\mathcal{E}_\mathrm{O}'(15\degr)}{\cos 15\degr}\approx 				   
        \begin{cases}   
         \hfill \un[16]{m^2}  &\text{at}\: \hfill H=\un[400]{km} \\   
         \hfill \un[42]{m^2}  &\text{at}\: \hfill H=\un[700]{km} \\   
         \hfill \un[104]{m^2} &\text{at}\: \hfill H=\un[1000]{km} \\   
        \end{cases} \\   
        D_\mathrm{min}&\approx 2\sqrt{\frac{A_\mathrm{min}}{\pi}}\approx   
        \begin{cases}   
         \hfill \un[4.5]{m}   &\text{at}\: \hfill H=\un[400]{km} \\   
         \hfill \un[7.5]{m}   &\text{at}\: \hfill H=\un[700]{km} \\   
         \hfill \un[12]{m}    &\text{at}\: \hfill H=\un[1000]{km} \\   
        \end{cases}           
    \end{split}   
    \punto   
\end{equation}   
   
These results cannot be applied without accounting for the other sources of   
inefficiencies in a real optical system, so that the lower limit on
$D_\mathrm{min}$ might increase by a factor,   
as confirmed by Monte-Carlo simulations~\cite{bib:esaf,TheaThesis,PesceThesis}.   
   
%-------------------------------------------------------------------------------   
\subsection{Granularity and angular resolution}\label{sec:AngResEst}   
%-------------------------------------------------------------------------------   
   
Requiring a spatial granularity on Earth   
surface $\DD{L} \lesssim \un[1]{km}$,    
the required pixel granularity at different orbital heights can be fixed:   
   
\begin{equation}\label{eq:DeltaAlphaEst}   
    \Delta\alpha \approx \cfrac{\DD{L}}{H} \approx   
    \begin{cases}   
       \hfill 0.10\degr &\text{at}\: \hfill H=\un[400]{km} \\   
       \hfill 0.06\degr &\text{at}\: \hfill H=\un[700]{km} \\   
       \hfill 0.04\degr &\text{at}\: \hfill H=\un[1000]{km} \\   
    \end{cases}      
    \punto   
\end{equation}   
   
Using the values of Table~\ref{tab:anglength} it turns out that the image of an
\EAS with $\theta=50\degr$ is about 17 pixels long and there are about 6 photons per pixel on average.    
   
The pixel size on the focal surface, from equation~\eqref{eq:pixels-1}, is
\mbox{$d\approx\un[5]{mm}$}, using the minimum diameter from
equation~\eqref{eq:ADest}, and $\Fnumb\sim 0.5$. 
This value of $d$ does not depend on $H$ since $D_\mathrm{min}\propto H$.   
   
It turns out that, in the reference conditions of
Table~\ref{tab:ReferenceConditions}, 
the angular resolution perpendicular and parallel to the line-of-sight are both    
\mbox{$\sim 0.5\degr$}. The total angular resolution is then     
\mbox{$\DD{\beta}_\mathrm{tot}=\sqrt{\DD{\beta_\perp}^2 + \DD{\beta_\parallel}^2}\lesssim 1\degr$}.

%-------------------------------------------------------------------------------   
\subsection{The Random Background}   
\label{sec:NightGlowEst}   
%-------------------------------------------------------------------------------   

The total random background rate intercepted on the whole entrance pupil by the telescope, in the reference conditions of Table~\ref{tab:ReferenceConditions},    
with $\Gmax=20\degr$ is given in Table~\ref{tab:RB}, together with the   
corresponding total number of pixels and the total random background rate per
pixel (with a total efficiency $\varepsilon_\mathrm{tot} \approx 0.1 $).   

\begin{table}[htb]   
\begin{center}   
\begin{tabular}{c|c|c|c} \hline   
H &   $N_\mathrm{pix}$   &  $\nu_\mathrm{tot}^\mathrm{b}$  & $\nu_\mathrm{pix}^\mathrm{b}$  \\   
 $[\text{km}]$ & & [THz] & [MHz] \\ \hline \hline   
400    &  \sci{1.2}{5}       &  2.8                      & 2.3 \\   
700    &  \sci{3.5}{5}       &  7.6                      & 2.2 \\   
1000   &  \sci{7.8}{5}       &  19.0                     & 2.4 \\ \hline   
\end{tabular}   
\end{center}   
\caption{Number of pixels, total random background rate and background rate per
  pixel, in the reference conditions of Table~\ref{tab:ReferenceConditions}. The values of $A_\mathrm{EP}$ and $\DD{\alpha}$ from equations~\eqref{eq:ADest}    
and~\eqref{eq:DeltaAlphaEst}, respectively, are used.}   
\label{tab:RB}   
\end{table}

The results show that there is  
one order of magnitude more background than signal photons    
superimposed on the typical \EAS (all space-time length) and   
roughly the same number of signal and random background photons near the \EAS maximum.

%-------------------------------------------------------------------------------   
\section{Estimation of the performance of \SEUSO}\label{sec:SEUSO}   
%-------------------------------------------------------------------------------   
   
The \SEUSO mission~\cite{supereuso}    
was proposed to ESA in the framework of the Cosmic Vision program 2015-2025~\cite{cosmicvision}.   
It is an enlarged and improved free-flyer version of the former \EUSO mission~\cite{EUSO}, modifying several aspects and exploiting novel technologies.   
The design of the experimental telescope exploits the results of the detailed studies carried on during the phase A of   
EUSO~\cite{EUSORedBook} to redesign the telescope in such a way to fully satisfy the Scientific Requirements listed in Section~\ref{sec:SciReqs}.    
   
The estimation of the   
performance of \SEUSO is relative to \EUSO~\cite{EUSORedBook}.    
It safely relies on the \EUSO phase-A studies, based on full simulations,   
including the signal generation and transport, the telescope response as well as data analysis~\cite{bib:esaf}. The \EUSO   
performance is rescaled to the new telescope on the basis of simple approximate scaling   
laws and/or consolidated expectations based on new technologies/techniques
developments. 
   
The optics efficiency can be improved by a factor $\approx 1.5$ with respect to \EUSO, by using a different, catadioptric, optical system with a slightly reduced   
FoV ($\Gmax=20\degr \div 25 \degr$). The instantaneous geometrical aperture is then recovered by means of higher altitude orbits.   
   
The photo-detection efficiency can realistically improve by a factor $\approx 3$
with respect to \EUSO, 
thanks to newly developed photo-sensors with a larger
quantum efficiency and a better flexibility in the design, which will allow to
improve the filling factor of the focal surface.   
   
As the entrance pupil size is the   
only parameter affecting the performance of the experiment whose size can be
chosen, to within practical constraints, the telescope is designed to have the largest   
possible size compatible with a non-deployable photo-detector. In fact, while
large deployable optical systems are under    
development for many other applications~\cite{bib:depoptics}, the engineering of
a deployable photo-detector does not appear as a realistic option.
In order to cope with the very large optics required, a reflective
deployable optical system is preferred. A reflective optical system has also
the additional advantage to allow a smaller \Fnumb than a refractive
optical system, thus helping to limit the size of the   
focal surface. The $\Fnumb = 0.7$ is used. With all these assumption, the
increase of the entrance pupil size can give a factor $\approx 10$ improvement
in the efficiency of the optics.   

The three above factors give an overall $\approx45$ factor improvement in the photon collection capability.    
   
The satellite orbit is an elliptical one, with the apogee in the range
\mbox{$r_\mathrm{a}\approx(800\div1200)\,\text{km}$} and the perigee as low as
possible, compatibly with    
the constraints from the space mission, including atmospheric drag. 
The perigee is in the range \mbox{$r_\mathrm{p}\approx(600\div1000)\,\text{km}$}.    
With a perigee twice   
larger than the \EUSO one, there is still a factor $\approx 45/4=11.25 $ improvement with
respect to \EUSO in the overall photon detection capability. Since the \EUSO efficiency as a function of the \EAS energy
reached a plateau at about    
$E\approx\un[\sci{2}{20}]{eV}$,
(for \EAS in the zenith angle range $30\degr < \theta < 70\degr$),
the energy threshold of \SEUSO will be at $E\approx\un[\sci{2}{19}]{eV}$.
Moreover, for \EAS with zenith angle $\theta > 60\degr$, the \EUSO efficiency as a function of the \EAS energy
reached a plateau at $E\approx\un[\sci{1}{20}]{eV}$, so that
\SEUSO would have an energy threshold of $E\approx\un[\sci{1}{19}]{eV}$.   
   
The \EAS triggering and reconstruction efficiency might be
improved as well, by exploiting the larger number of photons with respect to
\EUSO, but this is not taken into account in the present analysis.    
   
The result coming from all these assumptions lead to the parameters   
summarized in Table~\ref{tab:SEUSOparameters}.

\begin{table*}[htb]   
	\centering   
	\begin{tabular}{c|c|c}\hline   
\multicolumn{3}{c}{Satellite and Orbit} \\ \hline    
Orbit perigee & $r_\mathrm{p}\approx\un[800]{km}$  & $r_\mathrm{p}\approx(600\div1000)\,\text{km}$ \\   
Orbit apogee  & $r_\mathrm{a}\approx\un[1100]{km}$ & $r_\mathrm{a}\approx(800\div1200)\,\text{km}$ \\   
Orbit inclination & $i\approx 50\degr \div 60\degr$ & preferred \\   
Orbital period & $T\approx \un[100]{min}$ & \\   
Pointing and pointing accuracy  & nadir to within 3\degr & must be known offline to within 0.5\degr \\    
Lifetime & $>5$ years & 10 years goal \\ \hline \hline   
\multicolumn{3}{c}{Telescope} \\ \hline    
Type & deployable catadioptric system & \\   
Main mirror diameter & $D_\mathrm{M} = \un[11]{m}$  & \\   
Entrance pupil diameter & $D_\mathrm{EP} = \un[7]{m}$ & \\   
Field of View & $\Gmax=25\degr$ & half-angle \\   
Angular granularity & $\DD\alpha\approx 0.04\degr$ & \\   
Linear granularity & $\DD L \approx \un[0.7]{km}$ at Earth & average \\   
Optical triggering efficacy & $\mathcal{E}^\prime_\mathrm{O} \gtrsim \un[108]{m^2}$ & average requirement \\   
f-number & $\Fnumb\approx 0.7$ & current design (goal: $\Fnumb\approx 0.6$) \\   
Focal surface diameter & $D_\mathrm{PD} = \un[4]{m}$ & \\   
Photo-detection efficiency & $\PDE \gtrsim 0.25$ & average requirement \\   
Number of pixels & $N_\mathrm{pix}\approx \sci{1.2}{6}$ & \\   
Pixel size & $\delta\approx\un[4]{mm}$ & to be optimized \\ \hline \hline   
\multicolumn{3}{c}{Performance} \\ \hline    
Energy threshold & $E_\mathrm{th}\approx\un[\sci{1}{19}]{eV}$ & \\    
Instantaneous geometrical aperture & $\GeoAperture\approx \un[\sci{2}{6}]{km^2}\un{sr}$ & \\   
Statistical error on the energy & $\DD{E}/E \approx 0.1$ at $E=\un[\sci{1}{19}]{eV}$ & requirement \\   
Angular resolution & $\DD\beta \lesssim 3 \degr$ & depends on the \EAS direction \\   
Observational duty cycle & $\eta_\mathrm{o} \approx 0.1 \div 0.2$ & dedicated measurements required \\ \hline    
   
	\end{tabular}   
	\label{tab:SEUSOparameters}   
\caption{Main baseline parameters of the \SEUSO mission.}   
\end{table*}


\begin{thebibliography}{10}   
   


\bibitem{Abraham:2008ru}   
J.~Abraham et~al.,   
\newblock { Phys. Rev. Lett.} 101 (2008) 061101.   
   
\bibitem{Zavrtanik}   
D.~Zavrtanik,   
\newblock { Contemp. Phys.} 51 (2010) 513--529.    
   
\bibitem{Watson:2008zzb}   
A.A.~Watson,   
\newblock {Nucl. Instrum. Meth.} A 588 (2008) 221--226.   
   
\bibitem{Matthews:2011zz}
\newblock{J.N.~Matthews et~al., Nucl.\ Phys.\ Proc.\ Suppl.\  {212-213} (2011) 79.}
\newblock{H.~Tokuno et~al.,   J.\ Phys.\ Conf.\ Ser.\  293 (2011) 012035.}
  %``First results from the Telescope Array,''
  %``The status of the Telescope Array experiment,''

\bibitem{Abraham:2007si}   
J.~Abraham et~al.   
\newblock{{Astropart. Phys.} 29:188--204, 2008;}
\newblock{Astropart.\ Phys.\  {31} (2009) 399;}   
\newblock{ Phys.\ Rev.\  D {79} (2009) 102001;}   
\newblock{Astropart.\ Phys.\  {32} (2009) 89,   
  [Erratum-ibid.\  {33} (2010) 65];}   
\newblock{Phys.\ Lett.\  B {685} (2010) 239;}
\newblock{Phys.\ Rev.\ Lett.\  {104} (2010) 091101;}   
\newblock{Astropart.\ Phys.\  {33} (2010) 108.}  
   

   
\bibitem{Abreu:2010zzj}   
P.~Abreu {et al.},   
\newblock{Astropart.\ Phys.\  {34} (2010) 314--326;}
\newblock{Astropart.\ Phys.\  {34} (2011) 627--639;}
\newblock{JCAP {1106} (2011) 022.}


\bibitem{Abbasi:2007sv}
  R.U.~Abbasi {et al.},
  %``First observation of the Greisen-Zatsepin-Kuzmin suppression,''
  Phys.\ Rev.\ Lett.\  100 (2008)  101101.

\bibitem{Takeda:1998ps}
  M.~Takeda {et al.},
  %``Extension of the cosmic ray energy spectrum beyond the predicted
  %Greisen-Zatsepin-Kuz'min cutoff,''
  Phys.\ Rev.\ Lett.\  81 (1998) 1163.
   
   
\bibitem{Pallavicini:2008wy}   
  M.~Pallavicini, R.~Pesce, A.~Petrolini and A.~Thea,   
  The observation of Extensive Air Showers from Space,   
  arXiv:0810.5711 [astro-ph].   
   
\bibitem{arisaka}   
K.~Arisaka,   
\newblock Optimization of an {OWL-Airwatch} Optics and   
  Photo-Detectors (1999), \url{www.ge.infn.it/euso/docs/arisaka.pdf}.   
   
\bibitem{bib:esaf}   
  C.~Berat {et al.},   
  %``ESAF: Full Simulation of Space-Based Extensive Air Showers Detectors,''   
  Astropart.\ Phys.\  {33} (2010) 221.   

   
\bibitem{EUSO}   
J.~Adams,   
\newblock  {Nucl. Phys. Proc. Suppl.} 134 (2004) 15--22.

\bibitem{bib:petrolini1}   
A. Petrolini, Nucl. Instr. Meth. A588 (2008) 201--206.   
   
\bibitem{spacepart06}   
A.~Thea et~al,   
\newblock {Nucl. Phys. B-Proc. Suppl.} 166 (2006) 223--228.   

\bibitem{petrosanta}   
A. Santangelo and A. Petrolini,   
New J. Phys. 11 (2009) 065010. 

\bibitem{linsley}   
  R.~Benson and J.~Linsley,   
  \textit{Satellite observation of cosmic ray air showers},   
Proc. 17th ICRC (1981).  
   
\bibitem{owl}   
  F.W.~Stecker et al.,   
  %``Observing the ultrahigh energy universe with OWL eyes,''   
  Nucl.\ Phys.\ B-Proc.\ Suppl.\  {136C} (2004) 433.   
   
\bibitem{tus}   
V.I. Abrashkin et~al.,   
\newblock {Int. J. Mod. Phys.}, A20 (2005) 6865--6868.

\bibitem{jemeuso}   
Y. Takahashi et al., New Journal of Physics, Vol. 11, N. 065009, 2009   
   
\bibitem{supereuso}   
A.~Santangelo, A.~Petrolini, et~al,   
\newblock {{S-EUSO}: a proposal for a space-based observatory for   
  {UHECP}},   
\newblock Technical report, 2007,   
\newblock \url{www.ge.infn.it/euso/docs/S-EUSO\_Proposal.pdf}.   
    
   
\bibitem{petrolini2}   
A. Petrolini,   
Nucl. Instrum. Meth. A630 (2011) 131--135.   
   
\bibitem{cosmicvision}   
{ESA},   
\newblock {Cosmic Vision. Space Science for Europe 2015-2025}, 2005, \url{http://www.esa.int/esapub/br/br247/br247.pdf}.   
   
\bibitem{Albuquerque:2005nm}
  I.F.M.~Albuquerque and G.F.~Smoot,
  %``GZK cutoff distortion due to the energy error distribution shape,''
  Astropart.\ Phys.\  {\bf 25} (2006) 375.
   
\bibitem{EUSORedBook}   
The~EUSO Collaboration.   
\newblock {{EUSO: Report on the Phase A Study}},    
\newblock Internal note \EUSO-PI-REP-002 (2003), \url{http://www.ge.infn.it/euso/docs/EUSO-PI-REP-002.pdf}. 


%\bibitem{Allard:2008gj}
%  D.~Allard, N.~G.~Busca, G.~Decerprit, A.~V.~Olinto and E.~Parizot,
%  %``Implications of the cosmic ray spectrum for the mass composition at the
%  %highest energies,''
%  JCAP {\bf 0810} (2008) 033.
   


\bibitem{US-STD-76}   
US Standard Atmosphere http://modelweb.gsfc.nasa.gov/atmos/us\_standard.html.   
 
   
\bibitem{allen}   
A.N. Cox (ed.),   
\newblock {Allen's Astrophysical Quantities}.   
\newblock AIP Press, 1999.   
   
    
\bibitem{gaisser-hillas}   
T.~K. Gaisser and A.~M. Hillas,   
\newblock {Proc. 15th ICRC} 8 (1977) 353.   
   
   
\bibitem{Pryke}   
C.~Pryke,   
\newblock {Astropart. Phys.} 14 (2001) 319--328.   
   
   
\bibitem{stanev}   
T.~Stanev,   
\newblock {High Energy Cosmic Rays}.   
\newblock Springer, 2004.   
   
\bibitem{Bergmann:2006yz}   
  T.~Bergmann {et al.},   
  %``One-dimensional hybrid approach to extensive air shower simulation,''   
  Astropart.\ Phys.\  {26} (2007) 420.   
   
   
\bibitem{baby}   
O.~Catalano et~al.,   
\newblock {Nucl. Instrum. Methods A} 480 (2002) 547--554.   
   
   
\bibitem{kakimoto}   
F.~Kakimoto et~al.,   
\newblock {Nucl. Instrum. Meth.} A 372 (1996) 527--533.   
   
\bibitem{Nagano:2004am}   
M.~Nagano et~al.,   
\newblock {Astropart. Phys.} 22 (2004) 235--248.   

\bibitem{bib:flash}
R.~Abbasi et~al., Astropart. Phys. 29 (2008) 77.

\bibitem{Ave:2011ub}
  M.~Ave et al.,
  Nucl.\ Phys.\ Proc.\ Suppl.\  {\bf 212-213 } (2011)  356-361.

\bibitem{Bucholtz:1995}   
A.~Bucholtz,   
\newblock {Appl. Optics} 34(15) (1995) 2765--2773.   

\bibitem{Young:1994}   
A.~T. Young,   
\newblock {Appl. Optics} 33(6) (1994) 1108--1110. 

\bibitem{bib:Grieder}
P.~Grieder, \emph{Cosmic rays at Earth}, Elsevier, 2001.

\bibitem{Sokolsky:2003vj}
  P.~Sokolsky and J.~Krizmanic,
  %``Effect of clouds on apertures of space - based air fluorescence
  %detectors,''
  Astropart.\ Phys.\  20 (2004) 391.

\bibitem{AbuZayyad:2003af}
  T.~Abu-Zayyad, C.C.H.~Jui and E.C.~Loh,
  %``The effect of clouds on air showers observation from space,''
  Astropart.\ Phys.\  21 (2004) 163.


\bibitem{wertz}   
J.~Wertz and W.~Larson (eds.),   
\newblock {Space missions analysis and design},   
\newblock Microcosm Press, 1999. 

\bibitem{mazzinghi}   
P.~Mazzinghi et~al,   
\newblock {Physics and Astrophysics in Space} {(Frascati Physics Series)} vol.   
  xxxvii (2004) 437-444.   

%\bibitem{bib:bass}
%M.~Bass~(ed.), {Handbook of Optics}, McGraw-Hill, 2001.
     
%\bibitem{Piccioli:2003vq}   
%A.~Piccioli et~al.,   
%\newblock {Nucl. Instrum. Meth.} A 504 (2003) 294--297.   
   
%\bibitem{Buzhan:2003ur}   
%P.~Buzhan et~al.,   
%\newblock {Nucl. Instrum. Meth.} A 504 (2003) 48--52.   
   
%\bibitem{Golovin:2004jt}   
%V.~Golovin and V.~Savelev.,   
%\newblock {Nucl. Instrum. Meth.} A 518 (2004) 560--564.   

\bibitem{Nakamura:2010zzk}
\newblock{ T.~Gys, Nucl. Instrum. Meth. A {595} (2008) 136. }
\newblock{ M.~Suyama and K.~Nakamura, \textit{Recent progress of photocathodes for PMTs}, \url{http://pos.sissa.it/archive/conferences/090/013/PD09_013.pdf};}
\newblock{ K.~Nakamura, Y.~Hamana, Y.~Ishigami and T.~Matsui, Nucl.\ Instrum.\ Meth.\  A {\bf 623} (2010) 276.}
\newblock{ T.~Iijima, Nucl. Instrum. Meth. A {639} (2011) 137. }


\bibitem{ProcNDIP}
\newblock{ R.~Chechik and A.~Breskin, Nucl. Instrum. Meth. A {595} (2008) 116. }
\newblock{ J.~Haba, Nucl. Instrum. Meth. A {595} (2008) 154. }
\newblock {Nucl. Instrum. Meth.} A 610 (2009) 1--450;   
\newblock{ \textit{New Developments In Photodetection}, NDIP11, Proceedings of the 6th International Conference on New Developments in Photodetection.}
\newblock{ S.~Korpar, Nucl. Instrum. Meth. A {639} (2011) 88. }
\newblock{ S.~Dalla~Torre, Nucl. Instrum. Meth. A {639} (2011) 111. }
   
\bibitem{TheaThesis}   
A.~Thea,   
\newblock {Observation of UHECR from space.}   
\newblock PhD thesis, Universit\`{a} degli Studi di Genova, 2006, \url{http://www.ge.infn.it/euso/docs/AThea\_PhDThesis.pdf}.   
   
\bibitem{PesceThesis}   
R.~Pesce,   
\newblock {The observation of \UHECP on the Earth and from space},   
\newblock PhD thesis, Universit\`{a} degli Studi di Genova, 2008, \url{http://www.ge.infn.it/euso/docs/RPesce\_PhDThesis.pdf}.

\bibitem{KS}   
M.G.~Kendall, A.~Stuart, J.K.~Ord and S.~Arnold, \textit{Kendall's advanced theory of statistics}, Griffin, 2004.

\bibitem{Berat}   
C.~Berat et~al.,   
\newblock Proc. 28th ICRC (2003) 927--930.

\bibitem{bib:petroLSF}
A.~Petrolini, \textit{Least-squares fit to a straight line when each variable contains all equal errors}, arXiv:1104.3132 [physics.data-an].   

\bibitem{Sommers:1995dm}   
P.~Sommers,   
\newblock {Astropart. Phys.} 3 (1995) 349--360.   
   
\bibitem{Baltrusaitis:1985mx}   
R.~M. Baltrusaitis et~al.,   
\newblock {Nucl. Instrum. Meth.} A 240 (1985) 410--428.   
   
  
\bibitem{bib:UVng}
R.R.~Meier, Space Science Reviews
58(1) (1991) 1--185.

\bibitem{bib:tle}
U.S.~Inan et al., Geophys. Res.
Lett. 18 (1991) 705-708.

\bibitem{bib:tatianaflashes}
G.K.~Garipov et al., Astropart. Phys.
24 (2005) 400--408.
   
\bibitem{nightglow}   
L.M. Barbier et~al.,   
\newblock {Astropart. Phys.} 22 (2005) 439--449.   
   
\bibitem{arizona}   
D.J. Knecht et~al.,   
\newblock {Adv. Space Res.}, 19(4) (1997) 627--630.   
   
\bibitem{tatiana}   
G.K. Garipov et~al.,   
\newblock {JETP Lett.}, 82(4) (2005) 185--187.   
   
\bibitem{bib:depoptics}   
A.~Zuccaro~Marchi et al., \textit{Technological Developments for Ultra-Lightweight, Large Aperture, Deployable Mirror for Space Telescopes}, International Conference on Space Optics, ICSO 2010.   
   
   
   
   
   
   
\end{thebibliography}
\end{document}